\documentclass[twocolumn]{aastex62}

\newcommand{\teff}{\mbox{T$_{\rm eff}$}}
\newcommand{\logg}{\mbox{log~{\it g}}}
\newcommand{\vmicro}{\mbox{$\xi_{\rm t}$}}
\newcommand{\kmsec}{\mbox{km~s$^{\rm -1}$}}

\newcommand{\x}{\mbox{$\Delta_{\tiny{\mathrm{F275W,F814W}}}$}}
\newcommand{\y}{\mbox{$\Delta_{\tiny{C~\mathrm{ F275W,F336W,F438W}}}$}}

\submitjournal{ApJ}

\shorttitle{Chemical abundances in the Type~II GCs NGC\,1261 and NGC\,6934}  
\shortauthors{A.\,F. Marino, et al.}

\begin{document}


\title{SPECTROSCOPY AND PHOTOMETRY OF THE LEAST-MASSIVE TYPE-II GLOBULAR CLUSTERS: NGC1261 AND NGC6934\footnote{Based on observations collected at the European Southern Observatory under ESO programme 0101.D-0109(A), and the NASA/ESA Hubble Space Telescope, obtained at the Space Telescope Science Institute, which is operated by AURA, Inc., under NASA contract NAS 5-26555.}}

\author[0000-0002-1276-5487]{A.\,F.\.Marino}
\affiliation{Istituto Nazionale di Astrofisica - Osservatorio Astrofisico di Arcetri, Largo Enrico Fermi, 5, Firenze, IT-50125}


\author[0000-0001-7506-930X]{A.\,P.\,Milone}
\affiliation{Dipartimento di Fisica e Astronomia ``Galileo Galilei'', Universit\`{a} di Padova, Vicolo dell'Osservatorio 3, I-35122, Padua, Italy}
\affiliation{Istituto Nazionale di Astrofisica - Osservatorio Astronomico di Padova, Vicolo dell'Osservatorio 5, IT-35122, Padua, Italy}

\author[0000-0002-7093-7355]{A.\,Renzini}
\affiliation{Istituto Nazionale di Astrofisica - Osservatorio Astronomico di Padova, Vicolo dell'Osservatorio 5, IT-35122, Padua, Italy}

\author{D. Yong}
\affiliation{Research School of Astronomy \& Astrophysics, Australian National University, Canberra, ACT 2611, Australia}

\author[0000-0002-5804-3682]{M.\,Asplund}
\affiliation{Max Planck Institute for Astrophysics, Karl-Schwarzschild-Str. 1, D-85741 Garching, Germany}

\author[0000-0001-7019-0649]{G.\,S.\,Da Costa}
\affiliation{Research School of Astronomy \& Astrophysics, Australian National University, Canberra, ACT 2611, Australia} 

\author[0000-0003-4624-9592]{H.\,Jerjen}
\affiliation{Research School of Astronomy \& Astrophysics, Australian National University, Canberra, ACT 2611, Australia} 

\author[0000-0002-7690-7683]{G.\,Cordoni}
\affiliation{Dipartimento di Fisica e Astronomia ``Galileo Galilei'', Universit\`{a} di Padova, Vicolo dell'Osservatorio 3, I-35122, Padua, Italy}

\author[0000-0003-1757-6666]{M.\,Carlos}
\affiliation{Dipartimento di Fisica e Astronomia ``Galileo Galilei'', Universit\`{a} di Padova, Vicolo dell'Osservatorio 3, I-35122, Padua, Italy}

\author[0000-0003-1713-0082]{E.\,Dondoglio} 
\affiliation{Dipartimento di Fisica e Astronomia ``Galileo Galilei'', Universit\`{a} di Padova, Vicolo dell'Osservatorio 3, I-35122, Padua, Italy}

\author[0000-0003-1713-0082]{E.\,P.\,Lagioia}
\affiliation{Dipartimento di Fisica e Astronomia ``Galileo Galilei'', Universit\`{a} di Padova, Vicolo dell'Osservatorio 3, I-35122, Padua, Italy}

\author[0000-0002-7690-7683]{S.\,Jang}
\affiliation{Dipartimento di Fisica e Astronomia ``Galileo Galilei'', Universit\`{a} di Padova, Vicolo dell'Osservatorio 3, I-35122, Padua, Italy}

\author[0000-0002-1128-098X]{M.\,Tailo}
\affiliation{Dipartimento di Fisica e Astronomia ``Galileo Galilei'', Universit\`{a} di Padova, Vicolo dell'Osservatorio 3, I-35122, Padua, Italy}

\correspondingauthor{A.\ F.\,Marino}
\email{anna.marino@inaf.it}

\begin{abstract}
Recent work has revealed two classes of Globular Clusters (GCs), dubbed Type-I and Type-II.   Type-II GCs are characterized by a blue- and a red- red giant branch  composed of stars with different metallicities, often coupled with distinct abundances in the {\it  slow}-neutron capture elements ($s$ elements).

Here we continue the chemical tagging of Type-II GCs by adding the two least-massive clusters of this class, NGC~1261 and NGC~6934.
Based on both spectroscopy and photometry, we find that red stars in NGC~1261 to be slightly enhanced in [Fe/H]  by $\sim$0.1~dex and confirm that red stars of NGC~6934 are enhanced in iron by $\sim$0.2~dex. Neither NGC~1261 nor NGC~6934 show internal variations in the $s$ elements, which suggests a GC mass threshold for  the occurrence of $s$-process enrichment.

We found a significant correlation between the additional Fe locked in the red stars of Type-II GCs and the present-day mass of the cluster. Nevertheless, most Type~II GCs retained a small fraction of Fe produced by SNe~II, lower than the 2\%; NGC~6273, M~54 and $\omega$~Centauri are remarkable exceptions. 

In the appendix, we infer for the first time chemical abundances of Lanthanum, assumed as representative of the $s$ elements, in
M~54, the GC located in the nucleus of the Sagittarius dwarf galaxy. Red-sequence stars are marginally enhanced in [La/Fe] by 0.10$\pm$0.06 dex, in contrast with the large [La/Fe] spread of most Type II GCs. We suggest that different processes are responsible for the enrichment in iron and $s$-elements in Type-II GCs.

\end{abstract}

\keywords{globular clusters: individual (NGC\,1261, NGC\,6934) --- chemical abundances -- Population II -- Hertzsprung-Russell diagram } 

\section{Introduction}\label{sec:intro}

Based on data collected through the {\it Hubble Space Telescope} ($HST$)
photometric survey of Milky Way globular clusters (GCs) we have
established a powerful photometric diagram to investigate multiple stellar populations,
 i.e. a combination of four $HST$ bands
(F275W, F336W, F438W and F814W) allowing the construction of multiple
color plots now nicknamed {\it Chromosome Maps} \citep[ChM,][]{Mil15, Mil17}.
The position of stars in these diagrams is especially sensitive to the abundance
of C, N and O elements via the molecules that they form
\citep[e.g.\,][]{Mar08}, as well as to  helium and the overall metallicity
\citep{Mil15, Mar19}. 
One of the most exciting discovery from ChM analysis is that two different
classes of GCs are hosted in the Milky Way.
A first class includes the majority of GCs, with their variations in
the elements involved in the hot-H burning, with the typical Na-O,
C-N, and sometimes Mg-Al, anticorrelations, and homogeneous abundances
in heavy elements. The ChM of these clusters is populated by two main
stellar groups: (1) one corresponding to the O-enhanced population, also
displaying normal Na, N, He for their metallicity (hereafter 1P), and (2) the
other one being depleted in O, C and enhanced in Na, N, He (hereafter 2P)
\citep{Mil15, Mar19}. 
We usually refer to this class of objects as Type~I GCs.

The second class of globulars represents $\sim$17\% of the GCs analysed
in the $HST$ survey. Their most distinctive feature is the presence of
``more than one sequence'' on the ChM. In addition to the ChM sequence made of 1P and 2P stars,
common to all Milky Way GCs, this class displays an additional
ChM sequence running on the red side of the main map. 

Many of these objects
have been spectroscopically classified as {\it anomalous} GCs, as, at
odds with a typical cluster, host stellar populations with different
metallicity (Z) and, often, different abundances in the neutron-capture
elements produced via slow-neutron capture reactions
\citep[$s$-elements, e.g.\,][]{Mar09, Mar11, Mar15, Y&G08,
  Yong14, Johnson2015a, Johnson2017a}. 
  Metallicity variations are associated with variations in [Fe/H], variations in the overall C$+$N$+$O abundance, or both \citep[e.g.][]{yong2009a, Mar12_Ome}. 

  The color-magnitude diagrams (CMDs) of these clusters show multiple sub-giant branches \citep[SGBs,][]{Mil08, Mar12_M22} and red-giant branches \citep[RGBs,][]{Mar15, Mar19}. Stars on the red RGB also define distinct sequences along the asymptotic giant branch and the horizontal branch \citep[e.g.][]{dondoglio2021a, lagioia2021a}. 
We dubbed this newly-discovered class of objects Type~II GCs.

The interest in Type~II GCs comes from the fact that
variations in the overall metallicity were considered a
characteristic of more massive stellar systems, such as galaxies, capable of retaining supernovae ejecta in contrast to ordinary
GCs. In this context, $\omega$~Centauri, with its major internal variations in
metals, has been always considered as the surviving nucleus of a
disrupted dwarf galaxy 
 \citep[e.g.\,][]{B&F03}. 
Evidence of tidal debris from this massive GC reinforces
the hypothesis that $\omega$~Centauri is indeed a dwarf galaxy remnant
\citep{Majewski12, Ibata19}.
On the other hand, it may well be that most GCs formed inside dwarfs, most of which later dissolved due to interactions within the growing Milky Way. Indeed, it is now widely accepted that many GCs are associated with stellar streams \citep[e.g.][]{massari2019a} and that  half of the known Type II GCs appear clustered in a distinct region of the integral of motions space, thus suggesting a common progenitor galaxy \citep[][]{milone2020a}.

Beside having internal variations in Fe, $\omega$~Centauri displays a clear enhancement in $s$-elements as metallicity increases
\citep[e.g.\,][]{N&DaC95, J&P10, Mar11b}, making the similarity in the
chemical pattern with the {\it anomalous} GCs remarkable
\citep[][]{DaC&Mar, Mar15}. 
Noticeably, one Type~II GC, NGC\,6715 (M\,54), lies in the central region of
the Sagittarius dwarf galaxy \citep{Bellazzini08}; and two others,
NGC\,1851 \citep{Olszewski09, Kuzma18} and M\,2 \citep{Kuzma16}, are
surrounded by extended envelopes.
The stellar halo surrounding NGC\,1851 has a chemistry compatible with
a galaxy remnant \citep{Mar14}.


As previously discussed, the ChM of all the known
{\it anomalous} GCs, e.g. those with metallicity variations, can be classified as a Type~II map, with the
Fe-richer stars populating the red additional patterns on the map
\citep[see][]{Mar19}.
We are conducting chemical tagging, based on high-resolution
spectroscopy, to infer the occurrence and size of the chemical
abundance variations in the known Type~II GCs. So far, we have
chemical abundances measured on both the blue and red sequence of
the ChM for eight Type~II GCs.
In the present work, we analyse chemical abundances of two among the less
massive Type~II GCs known so far,
namely NGC\,1261 and NGC\,6934 \citep{Mil17}. 
NGC\,1261 has never been analysed in the context of the link between
the Type~II morphology of the ChM and internal variations in heavy
elements. Indeed, as discussed in \citet{Mil17}, the stars
distributing on the additional redder  sequence 
provide a minor contribution to the GC mass, and only ChM-guided
spectroscopic campaigns can pinpoint them. \citet{Kuzma18} recently
found a stellar 
envelope surrounding this GC, suggesting that 
 it might be the product of dynamical evolution of the
cluster. Indeed, this stellar structure has a much smaller size than
that observed around NGC\,1851.

On the other hand, four giants along the ChM of NGC\,6934, two on the
typical blue sequence, and two other on the redder one, have been
already analysed \citep{Mar18}
 founding that the two redder stars have higher
Fe abundances, by $\sim$0.15~dex, with respect to the other two
stars, making this Type~II GC a chemical {\it anomalous} GC. However,
 in contrast with the known Type\,II GCs, the two red stars do
not show any evidence of enhancements in the $s$-elements. This
difference may indicate a different chemical evolution of NGC\,6934
with respect to the bulk of the {\it anomalous} GCs. 
Nevertheless, given
the small sample size of \citet{Mar18}, the presence of $s$-enriched
stars cannot be definitively ruled out.

The outline of the paper is as follows: Section~\ref{sec:data} is a
description of the spectroscopic and photometric data-sets; the choice
of the adopted atmospheric parameters is discussed in
Section~\ref{sec:atm}; while our chemical abundance analysis is
outlined in Section~\ref{sec:abundances}; Sections~\ref{sec:abb6934},
 and \ref{sec:abb1261} 
describe the results for NGC~6934 and NGC~1261, 
 respectively. 
Finally, all our results are summarized and discussed in Section~\ref{sec:conclusions}. 

Although M\,54 is providing perhaps the clearest evidence for a possible link between {\it anomalous} GCs and nuclei of dwarf galaxies, as far as we
know, there is no analysis of $s$-element abundances in the stars with
different Fe in this context. In the appendix, we have filled this gap by
measuring first chemical abundances of lanthanum in giant stars whose
spectra were analysed by \citet{car10b}. Our aim is to investigate if,
similarly to other Type~II GCs, these stars show internal variations
in those species belonging to the neutron-capture element group.

\section{Data}\label{sec:data}

\subsection{The photometric dataset: the chromosome map of NGC\,1261
  and NGC\,6934 \label{sec:phot_data}} 

Photometric data used in this study come from the $HST$ UV Legacy
Survey to investigate multiple stellar populations in GCs \citep[GO-13297,][]{Piotto15}.
Details on the images analyzed and data reduction can be found in
\citet{Piotto15} and \citet{Mil17}.

\begin{figure*}
\includegraphics[width=18cm]{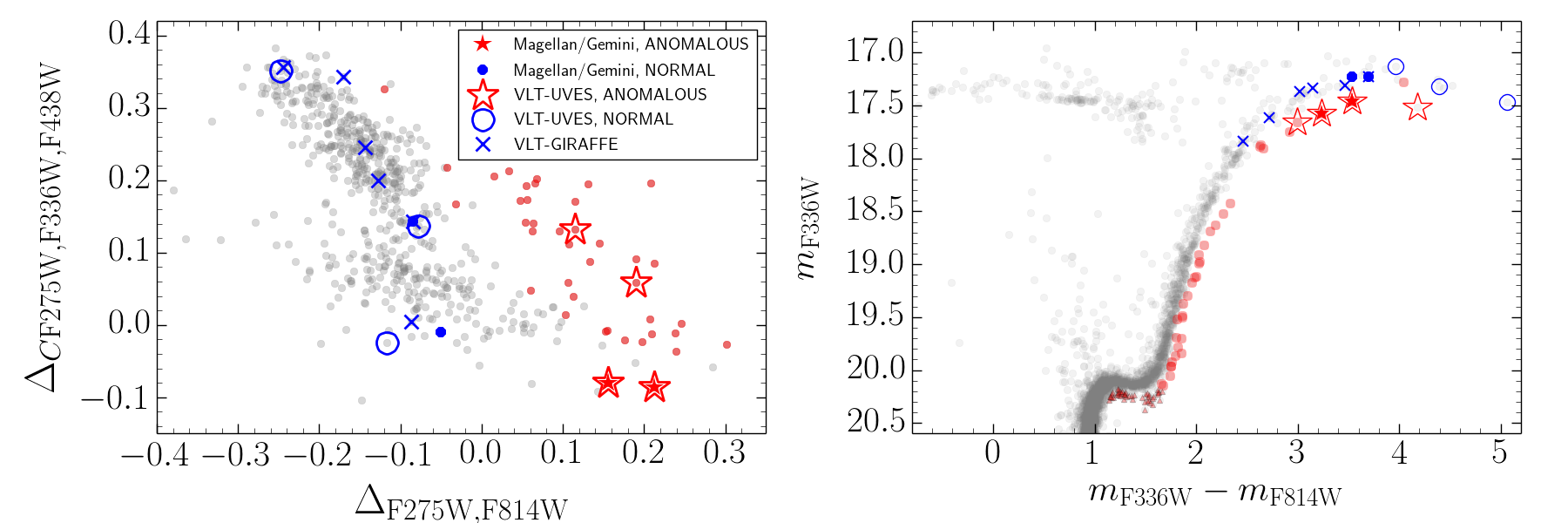}
\includegraphics[width=18cm]{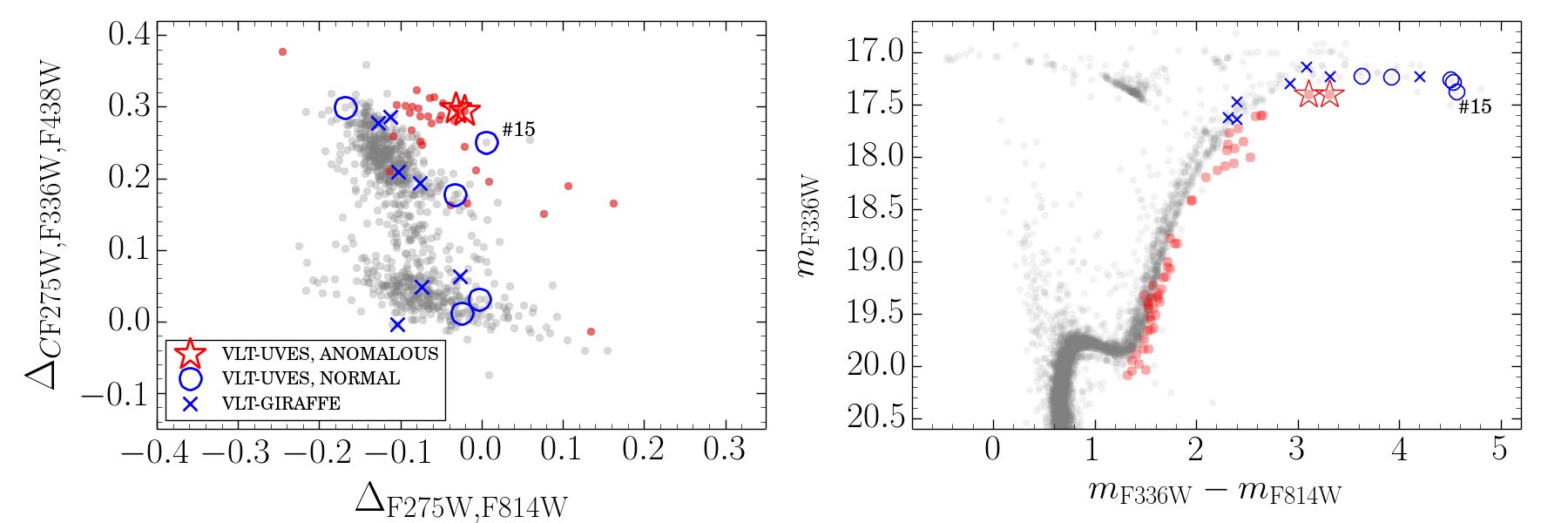}
     \caption{Chromosome maps (left panels) and $m_{\rm F336W}$ vs.\, $m_{\rm
         F336W}-m_{\rm F814W}$ CMDs (right panels) of the GCs
       NGC\,6934 (upper panels) and NGC\,1261 (lower panels). 
      The stars coloured in red have been selected as the stars defining the
      redder RGB stars on the CMDs. Open circles and star-like
      symbols represent blue and red RGB stars, respectively,
      observed with UVES; crosses are the GIRAFFE targets, all
      classified as blue RGB stars. For NGC\,6934 we also show
      stars observed with Magellan or Gemini telescope from \citet{Mar19}.}
       \label{fig:NGC6934photUVES}
  \end{figure*}

\citet{Mil17} analyzed the ChMs of 57 GCs, including NGC\,1261 and NGC\,6934,
and noticed some peculiarities (see their Figure~4) with respect to typical GC maps.  
The ChM of red giants in both clusters is reproduced in
Figure~\ref{fig:NGC6934photUVES} (left panels), 
 where we can clearly distinguish:
{\it (i)} the presence of 1P and 2P stars, as the stars located at
lower and higher \y\ values;
and {\it (ii)} two distinct sequences of stars represented with gray and red dots,
respectively. As discussed in Section~\ref{sec:intro}, the presence of
the  separate distribution of red stars on the ChM is a distinctive feature of the Type~II GCs. 

In the right panels of Figure~\ref{fig:NGC6934photUVES} we show the 
$m_{\rm F336W}$ vs.\, $m_{\rm F336W}-m_{\rm F814W}$ CMDs of the two
analysed clusters. The distinction between blue and red-RGB stars in
Type~II GCs 
 can also be made on the CMD, 
 either by using the 
$m_{\rm F336W}$ vs.\, $m_{\rm F336W}-m_{\rm F814W}$ CMD based on $HST$
filters, or the $U$ vs.\,$U-I$ CMD from ground-based photometry \citep{Mar19}.
Then, the blue and red-RGBs are generally observed to define two distinct ChMs
sequences. Among our spectroscopic targets, we note just one star in
NGC\,1261 (\#15), observed with UVES and classified as blue-RGB on the CMD, but located in a
position consistent with redder stars on the ChM (see
Figure~\ref{fig:NGC6934photUVES}). In the following this star is
treated as a blue RGB star, consistently with its location on the
$m_{\rm F336W}$ vs.\, $m_{\rm F336W}-m_{\rm F814W}$ CMD.

\subsection{The spectroscopic dataset}\label{sec:spec_data}

Spectroscopic data for NGC\,1261 and NGC\,6934 have been acquired using the FLAMES Ultraviolet and
Visual Echelle Spectrograph \citep[FLAMES-UVES,][]{Pasquini00} on the
European Southern Observatory (ESO) Very Large Telescope (VLT),
through the program 0101.D-0109(A). The observations were taken in the
standard RED580 setup, which has a wavelength coverage of
4726-6835~\AA\ and a resolution R$\sim$47,000 \citep{Dekker00}. 
Together with the UVES fibers, we used GIRAFFE fibers with the HR13 setup to observe a
sample of RGBs in the spectral range from $\sim$6122~\AA\ to
$\sim$6402~\AA\ at a resolving power R$\sim$22,000.

Spectra are based on 9$\times$2775s exposures for
NGC\,6934, and 7$\times$2775s exposures for NGC\,1261. 
UVES targets have been selected to be distributed on both the
blue and red sequences of the CMD and the ChM (see previous
Section). For NGC\,6934 we have observed three blue RGBs and four red RGBs
(two of which were also observed in \citet{Mar19}), our sample for
NGC\,1261 is composed of four blue and two red RGBs.

Data were reduced using
the ESO pipelines for FLAMES within the EsoReflex interface\footnote{\sf
  {https://www.eso.org/sci/software/esoreflex/}} \citep{Ballester00},
including bias subtraction, flat-field correction, 
wavelength calibration, sky subtraction and spectral rectification. 
Once individual spectra were reduced, the telluric
subtraction has been performed by using the ESO MOLECFIT tool 
\citep[][]{smette14, kausch14}.
Co-addition of the individual exposures, continuum normalisation, and
correction to laboratory wavelength was performed using IRAF
routines. 

Radial velocities (RVs) were derived using the IRAF@FXCOR task, which
cross-correlates the object spectrum with a template. For the template
we used a synthetic spectrum generated with the latest version of
MOOG\footnote{{\sf http://www.as.utexas.edu/~chris/moog.html}} 
\citep[version June 2014,][]{moog}. 
This spectrum was computed with a model stellar
atmosphere interpolated from the \citet{C&K04} grid,
adopting parameters (effective temperature (\teff), surface gravity (\logg), microturbulence (\vmicro), metallicity ([A/H])) of 4600~K, 2.5, 2.0~\kmsec, and $-$1.45~dex,
respectively. Observed RVs were then corrected to the heliocentric
system. 

Heliocentric RVs were used to establish cluster membership for our
targets. In the end, our sample of stars on the ChM is composed of 13 red giant stars in NGC\,6934
and 14 in NGC\,1261 (both GIRAFFE and UVES).
For the two clusters we measure average RVs of
$\langle$RV$\rangle$=$-408.6\pm1.2$~\kmsec\ (r.m.s.=4.2~\kmsec), and
$\langle$RV$\rangle$=$+71.1\pm1.1$~\kmsec\ (r.m.s.=3.8~\kmsec), for
NGC\,6934 and NGC\,1261, respectively. These values are in good agreement with those reported in the 2010 version of the \citet{harris1996a} catalog ($-$411.4$\pm$1.6 and 68.2$\pm$4.6 for NGC\,6934 and NGC\,1261, respectively).

%

\section{Model atmospheres}\label{sec:atm}

First estimates of the atmospheric parameters for the program stars
have been derived by taking advantage of our photometry, from both
$HST$ and ground-based facilities. 
The $m_{\rm F606W}$ and $m_{\rm F814W}$ mag from $HST$ have been converted to $V$
and $I$ \citep{jay08}, which we then used to estimate effective
temperatures from the color-temperature calibrations \citep{alonso99}, assuming
mean reddening values of E$(B-V)$=0.10 and  E$(B-V)$=0.01 for
NGC\,6934 and NGC\,1261, respectively \citep{Harris10}.
Surface gravities were then obtained from the $V$ magnitudes,
the photometric temperatures, bolometric corrections from
\citet{alonso99}, distance modulus of $(m-M)_{V}$=16.28 for
NGC\,6934 and $(m-M)_{V}$=16.09 for NGC\,1261 \citep{Harris10}. The
stellar mass has been fixed to 0.70~${\rm M_{\odot}}$.
Photometric temperatures and surface gravities are listed in
Table~\ref{tab:atm}.

The high resolution and the large spectral coverage of the UVES
spectra allowed us to have a fully-spectroscopic estimate of
 \teff\ and \logg\ from Fe lines.
Hence, we determined \teff\ by imposing the excitation potential (E.P.)
equilibrium of the Fe\,{\sc i} lines and gravity with the ionization
equilibrium between Fe\,{\sc i} and Fe\,{\sc ii} lines. Note that for
\logg\ we impose Fe\,{\sc ii} abundances that are
0.06-0.08~dex higher than the Fe\,{\sc i} ones to adjust for non-local
thermodynamic equilibrium (non-LTE) effects \citep{bergemann12, LB&A12, amarsi2016a}.
The microturbulent velocity was set to minimize any dependence on
Fe\,{\sc i} abundances as a function of equivalent widths (EWs). 
Spectroscopic parameters for stars observed with UVES are listed in
Table~\ref{tab:atm}. 

By comparing the spectroscopic and photometric \teff/\logg\ obtained
from the UVES spectra for NGC\,6934 we notice that spectroscopic
\teff\ values are higher by $+$52$\pm$8~K (r.m.s.=20~K). Surface
gravities are on the 
same scale with average differences 
$\Delta$\logg=\logg$_{(V-I)}-$\logg$_{\rm
  {~Fe~lines}}$=$-$0.05$\pm$0.05 (r.m.s.=0.12). 
Similar offsets have been obtained for NGC\,1261, for which we get
spectroscopic \teff\ higher by $+$36$\pm$9~K (r.m.s.=22~K), and \\
$\Delta$\logg=\logg$_{(V-I)}-$\logg$_{\rm
  {~Fe~lines}}$=$-$0.01$\pm$0.03 (r.m.s.=0.07). 
This comparison suggests that internal estimates of uncertainties
associated with our atmospheric parameters are $\sim$50~K for \teff,
and $\sim$0.15~dex for \logg. For \vmicro\ and metallicity we adopt
typical internal uncertainties of 0.20~\kmsec\ and 0.10~dex, respectively.

For GIRAFFE spectra, once temperatures and gravities have been fixed
from photometry, we derived metallicitities and microturbulences from
the Fe~{\sc i} lines as explained above. As a comparison,
Table~\ref{tab:atm} also lists the \vmicro\ values from the
\vmicro-\logg\ empirical relation obtained by \citet{Mar08}. 
For the GIRAFFE targets we assume the same uncertainties
associated with the stellar parameters that we used for UVES.

\section{Chemical abundance analysis}\label{sec:abundances}

Chemical abundances were derived from a local thermodynamic
equilibrium (LTE) analysis by using the spectral analysis code MOOG
\citep{moog}, and the alpha-enhanced Kurucz model atmospheres of 
\citet{C&K04}, whose parameters have been obtained as described in
Section~\ref{sec:atm}.  

For UVES spectra we infer chemical abundances for proton-capture
elements O, Na, Mg and Al, for $\alpha$ elements Si, Ca, Ti ({\sc i}
and {\sc ii}), for Sc ({\sc ii}), V, Cr ({\sc i} and {\sc ii}), Mn, Fe ({\sc i}
and {\sc ii}), Co, Ni, Cu, Zn, and neutron-capture elements Y ({\sc ii}), Zr ({\sc ii}), Ba ({\sc
  ii}), La ({\sc ii}), Ce ({\sc ii}), Pr ({\sc ii}), Nd ({\sc ii}), Eu
({\sc ii}).
The chemical abundances for all the elements, with the exceptions of O,
Al, Mn, Cu, Zr, Ba, La, Ce, Pr and Eu, for which spectral synthesis
has been employed, 
have been inferred from an EW-based analysis. 
A detailed description of all the spectral features analysed from UVES
data can be found in \citet{Mar19b}.
The LTE approach is justified by the fact that we are mostly interested in abundance differences among different populations of stars with similar stellar parameters.  Indeed, relative non-LTE abundance corrections are expected to be negligible for our purposes.

\begin{center}
  \begin{figure*}
\includegraphics[width=18cm]{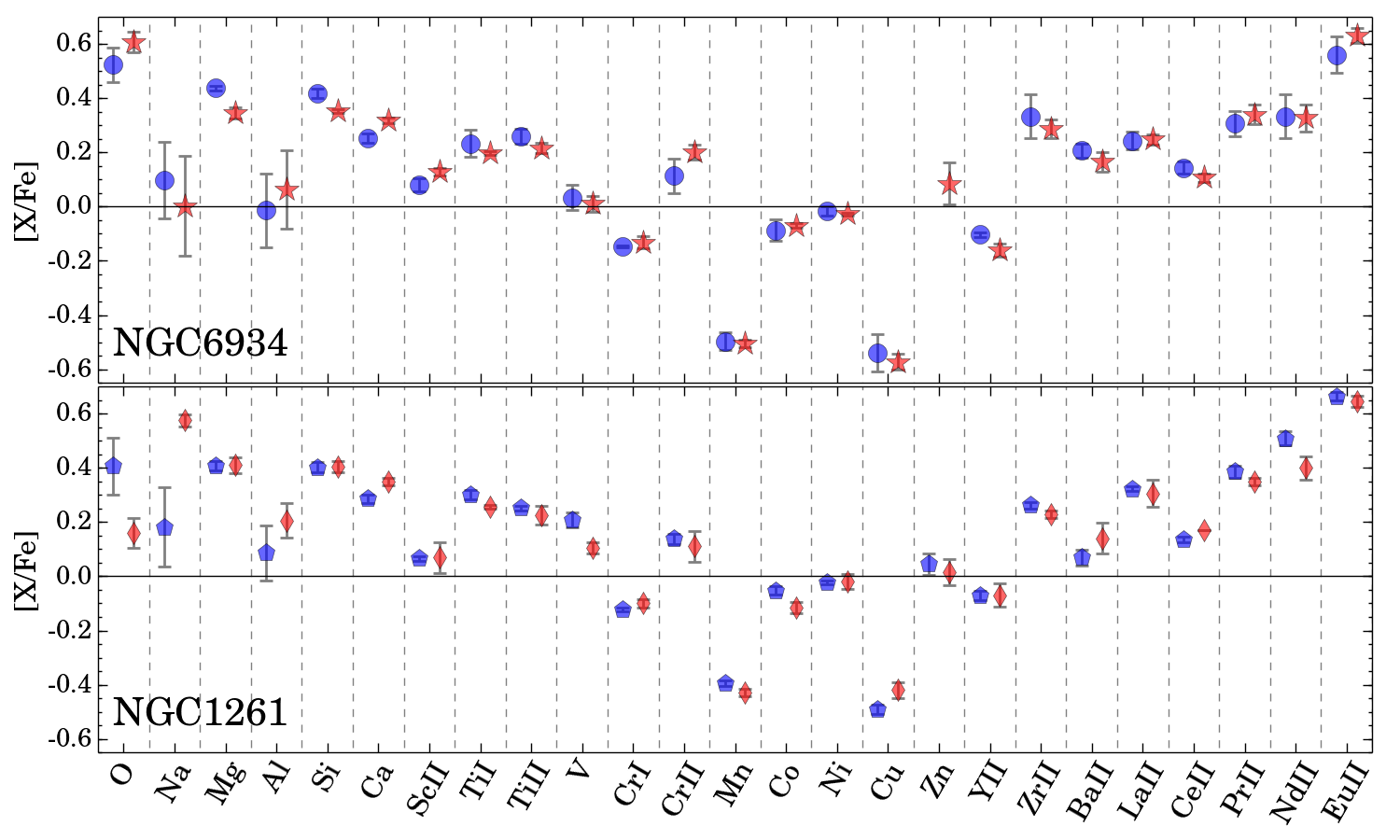}
     \caption{Summary of the abundance ratios results obtained from the UVES spectra. For each species, we plot the [X/Fe] relative abundances. Red symbols are used for stars located on the red sequence of the ChM, and filled blue symbols denote stars on the blue sequence.
      }
       \label{fig:box}
  \end{figure*}
 \end{center}

The chemical abundances obtained with UVES are listed in
Tables \ref{tab:LiToSc}--\ref{tab:CuToEu}, and the average for blue and red stars are plotted in
Figure~\ref{fig:box}, for both NGC~6934 and NGC~1261.
Internal uncertainties in chemical abundances due to the adopted model
atmospheres were estimated by varying the stellar parameters, one at a
time, by the amounts derived in Section~\ref{sec:atm}.
In addition to the contribution introduced by internal errors in
atmospheric parameters, we estimated the contribution due to the
finite S/N that affects the measurements of EWs and the spectral
synthesis.
For the species inferred from spectral synthesis we have varied the
continuum at the $\pm 1~\sigma$ level, and re-derived the chemical
abundances from each line.
For the elements with a number of available lines $\geq$5 we
assume the r.m.s. of the average abundance inferred from
individual spectral features divided by $\sqrt{N_{\rm lines}}$ as an estimate of the internal errors
associated with the errors in the EW measurements. For elements with only a 
few lines available, we follow the approach by \citet{norris10} and
\citet{yong13}: for each element, we replace the r.m.s. in
Tables~\ref{tab:LiToSc}--\ref{tab:CuToEu} with the maximum value. Then, we derive
max(r.m.s.)$/\sqrt{N_{\rm lines}}$. 
Since the EWs/continuum placement errors are random, the error 
associated to those elements with a larger number of lines,
e.g.\,Fe~{\sc i}, is lower.
The typical values obtained for each element are listed in 
Table~\ref{tab:err}. The total error is obtained by quadratically adding this
random error with the uncertainties introduced by atmospheric
parameters.

\section{The chemical composition of  NGC\,6934}\label{sec:abb6934}

The mean metallicity obtained from the UVES sample in NGC\,6934 is
[Fe/H]=$-$1.43$\pm$0.05~dex (rms=0.12~dex), which is consistent with
the two RGB stars observed with MIKE@Magellan \citep{Bernstein03} by
\citet{Mar18}.\footnote{As discussed in \citet{Mar18}, the GRACES@Gemini \citep{Chene14}
values are systematically higher by $\sim$0.20~dex.
The two stars belonging to the 
 red ChM sequence
observed in \citet{Mar18}, namely NGC\,6934-a1 and NGC\,6934-a2, 
have been re-observed with UVES and have [Fe/H]=$-$1.30$\pm$0.01~dex (\# 27) and
[Fe/H]=$-$1.35$\pm$0.01~dex (\# 47), respectively. The abundances reported in
\citet{Mar18} 
for these two stars are [Fe/H]=$-$1.10$\pm$0.02~dex for NGC\,6934-a1
(from GRACES spectra), and [Fe/H]=$-$1.35$\pm$0.01~dex for
NGC\,6934-a2 (from MIKE spectra, for the same star the GRACES value is
higher by 0.21~dex). 
Thus, we conclude that Fe abundances agree between the stars observed
with UVES and with MIKE, while there is a systematic offset by 0.20~dex with
GRACES spectra. }

The four UVES stars in the red 
 ChM sequence have
[Fe/H]=$-$1.34$\pm$0.02~dex (rms=0.03~dex), while for the three 
 blue-RGB stars we have inferred a mean iron abundance of 
[Fe/H]=$-$1.55$\pm$0.03~dex (rms=0.04~dex).
Quantitatively, we obtain $\Delta$[Fe/H]~{\sc i}=0.21$\pm$0.04~dex,
which is more than 5~$\sigma$ level.
These results suggest that the 
 red-RGB stars are enriched
in the overall metallicity, corroborating with higher statistical
significance, the results by \citet{Mar18}.

The presence of a difference in metallicity between the blue and red
stars on the ChM is further reinforced by the GIRAFFE targets, all
belonging to the blue ({\it normal}) population. For these stars we get
an average iron abundance of [Fe/H]=$-$1.62$\pm$0.02~dex
(rms=0.05~dex), within 2~$\sigma$ from the value obtained from UVES,
and with a comparable rms, which is much smaller than that obtained
for the whole UVES sample. 

The Na-O abundances plotted in Figure~\ref{fig:NaO} shows the typical
anticorrelation pattern observed in GCs. The stars both in the 
 red and in the 
  blue sequence have a range in Na.
This is the first detection of internal variations in
light elements in the {\it anomalous} population of NGC\,6934, as the
targets of \citet{Mar18} were selected to be located in the
lower part of the ChM with relatively low \y\ values, where we did not
expect such variations.
The middle panel of Figure~\ref{fig:NaO} displays a clear correlation
that is present between the \y\ axis of the ChM and the Na abundances,
as found for all the analysed Milky Way GCs \citep{Mar19}. 
Beside the higher \x\ values on the map, the red stars have on average higher O abundances, as shown in the
right panel of Figure~\ref{fig:NaO}. 

Figure~\ref{fig:box} displays a summary of the other chemical abundance
ratios, as the mean of the three blue-RGB (or {\it normal}) and the
four red-RGB (or {\it anomalous}) stars (top panel).
Chemical abundances relative to Fe for the two groups are
consistent within the observational errors for all the analysed
species. Large spreads are observed in Na and Al for both the {\it normal} and the
four {\it anomalous} stars.
Interestingly, there is no evidence for variations in the {\it
  neutron}-capture elements between the blue and red stars. 
This result is consistent with what was found in \citet{Mar18}, who
analysed two red stars in the low \y\ region of the map 
not finding any evidence of internal enrichment in the $s$-process
elements.
The authors, however, could not exclude $s$-process element
enrichment at higher \y\ level of the ChM, a region which they did
not explore. 
 The near identity of the abundances of most elements in the blue and red populations excludes that the red sequence was accreted from outside, but instead it was formed from material enriched in metallicity, possibly by supernovae from the blue-sequence population.

In the left panel of Figure~\ref{fig:LaFe} we plot the [La/Fe] of
NGC\,6934, assuming La as a 
representative of the $s$-elements, versus [Fe/H]. Clearly, while all
the four
red stars have higher Fe, they have similar
La, independently on their location along the \y\ axis of the ChM.
This result confirms that 
NGC\,6934 shows no evidence of additional internal variations
in $s$-process elements.  

\section{The chemical composition of  NGC\,1261}\label{sec:abb1261}

The average Fe abundance for the giants analysed with UVES
in NGC\,1261 is [Fe/H]=$-$1.28$\pm$0.02 (rms=0.05). 
By dividing the sample in {\it normal} and {\it anomalous} stars, we get mean Fe abundances of 
[Fe/H]=$-$1.30$\pm$0.01~dex (rms=0.03, five stars) and
[Fe/H]=$-$1.22$\pm$0.01~dex (two stars), respectively.
Although the difference between the two groups goes in the direction
observed for nearly all Type~II GCs, i.e. the {\it anomalous} stars
are Fe-richer, the small sample of stars (only two in the red
sequence) and the small difference ($<$0.10~dex) prevent us from drawing
definitive conclusions about a possible enrichment in Fe among the 
 red-sequence stars of this cluster by using spectroscopy alone.
Nevertheless, the slightly higher Fe of two stars, and their concurrent
position on the red side of the ChM 
 suggest a 
small Fe enrichment in these stars.

For the seven stars analysed with GIRAFFE we obtain an average Fe~{\sc i}
abundance of [Fe/H]=$-$1.35$\pm$0.02 (rms=0.04). These stars
all belong to the blue sequence 
 on the ChM. Hence, a comparison
with the UVES chemical abundances is more appropriate by
considering only the 
 blue sequence stars in that sample. The average Fe
 abundances of the two samples agree at a 2~$\sigma$ level. 

As what was displayed for NGC\,6934, the lower panel of
Figure~\ref{fig:box} shows the mean abundances obtained for the 
 blue-sequence and red-sequence 
 stars of NGC\,1261 from UVES spectra.
We note that the overall chemical pattern between the two clusters is
similar. 

\begin{figure*}
  \includegraphics[width=18cm]{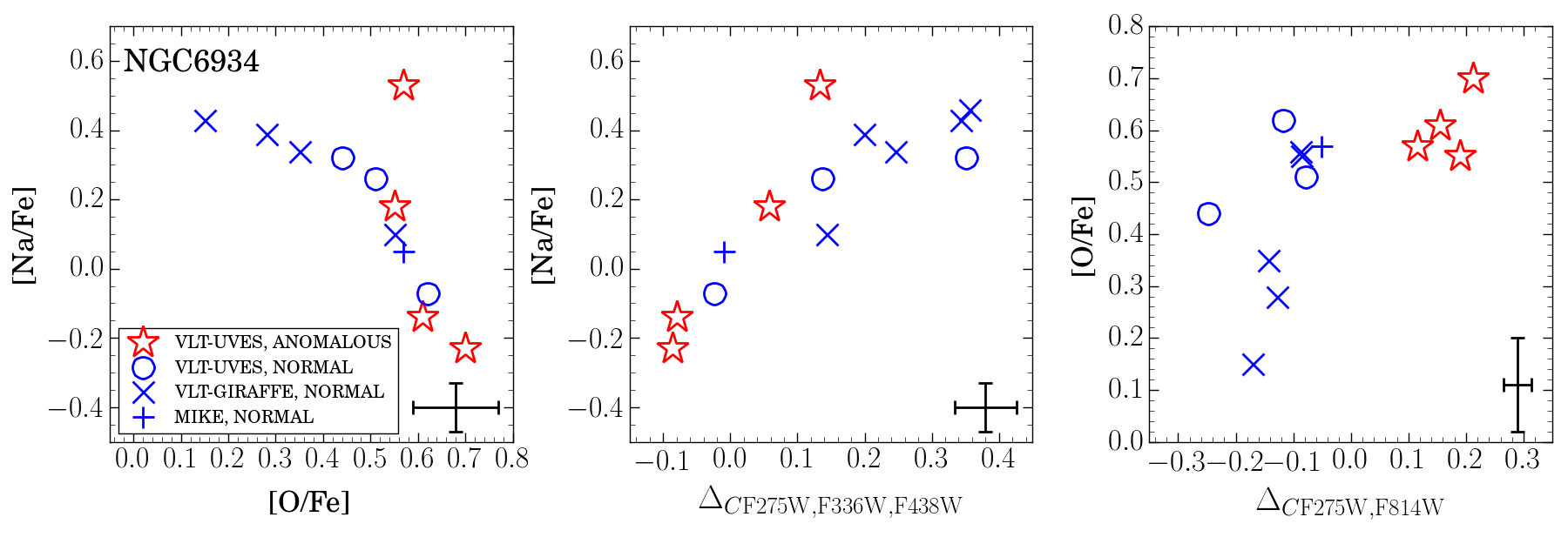}
  \includegraphics[width=18cm]{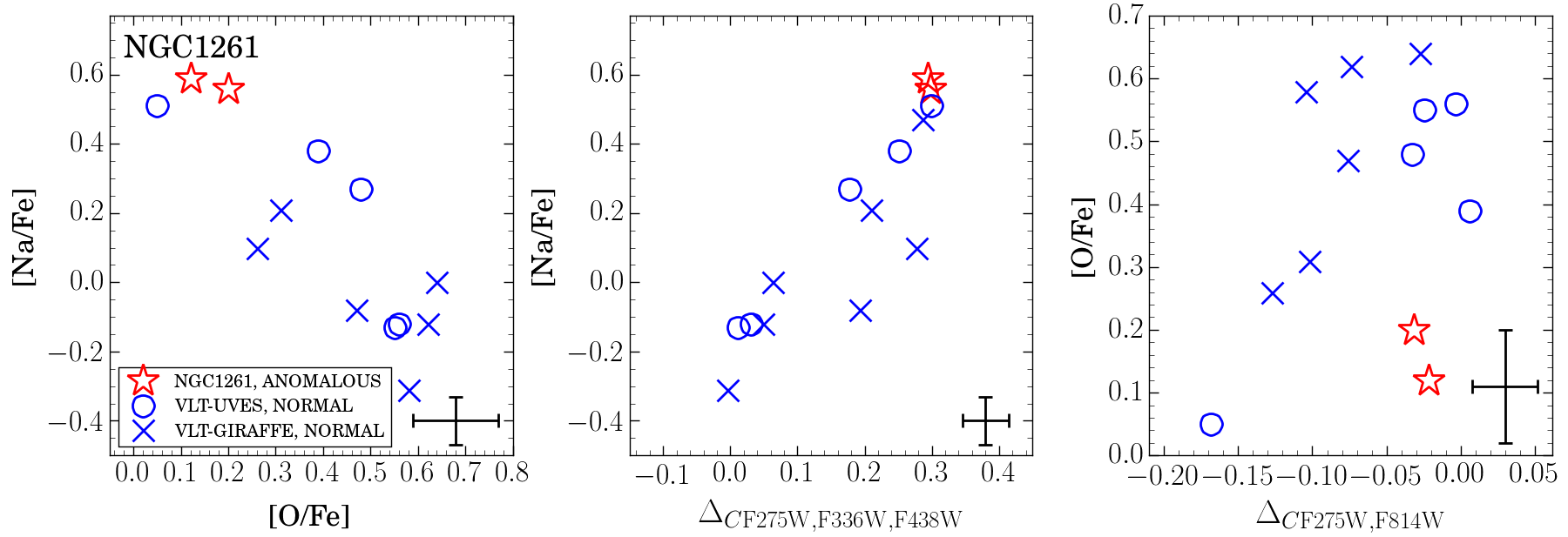}
     \caption{Na-O anticorrelation obtained for NGC~6934 (upper-left
       panel) and NGC~1261 (bottom-left panel). The middle and right
       panels represent the [Na/Fe] and [O/Fe] abundance as a function
       of the \y\ and \x\ axis of the ChM, respectively, for NGC~6934
       (upper panels) and NGC~1261 (bottom panels). Blue and red
       symbols represent the tagets on the blue and red sequence of
       the CMD and the ChM in both GCs.}
       \label{fig:NaO}
  \end{figure*}

Large spreads are observed in the light elements O, Na and
Al, while the Mg abundance range is narrow both among the {\it normal} and
the {\it anomalous} stars.
In the lower panels of Figure~\ref{fig:NaO} we show the Na-O
anticorrelation obtained from both the UVES and GIRAFFE data, as well as the Na
and O abundances as a function of \y\ and \x, respectively.
The two red stars also have the highest Na abundances, consistent 
with their higher \y\ values on the ChM. They are among the most 
O-depleted stars in the sample, not following the mild O-\x\ correlation defined by the blue stars.

\begin{center}
  \begin{figure*}
  \includegraphics[width=8.cm]{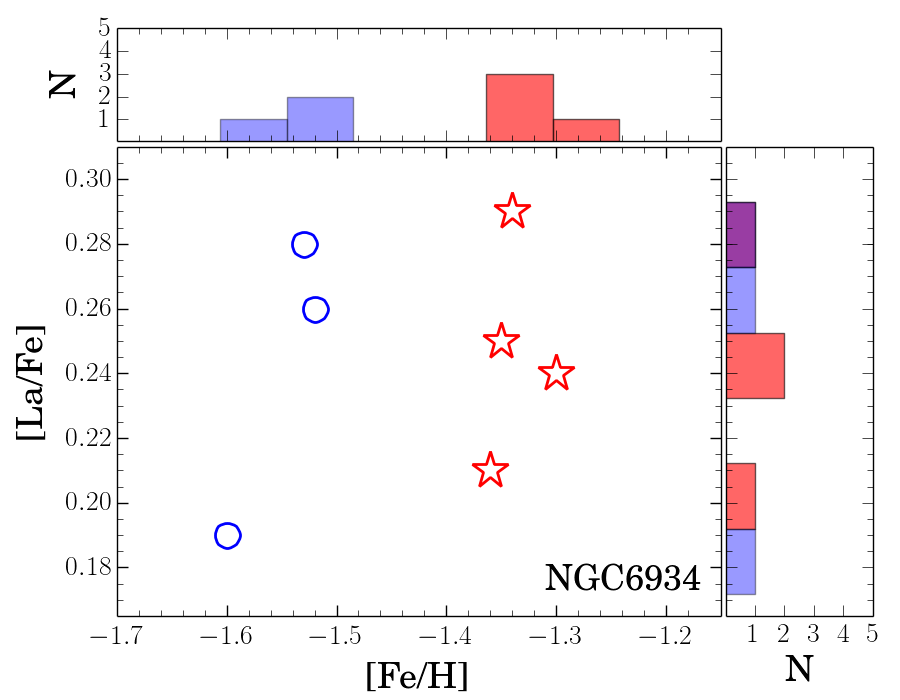}
  \includegraphics[width=8.cm]{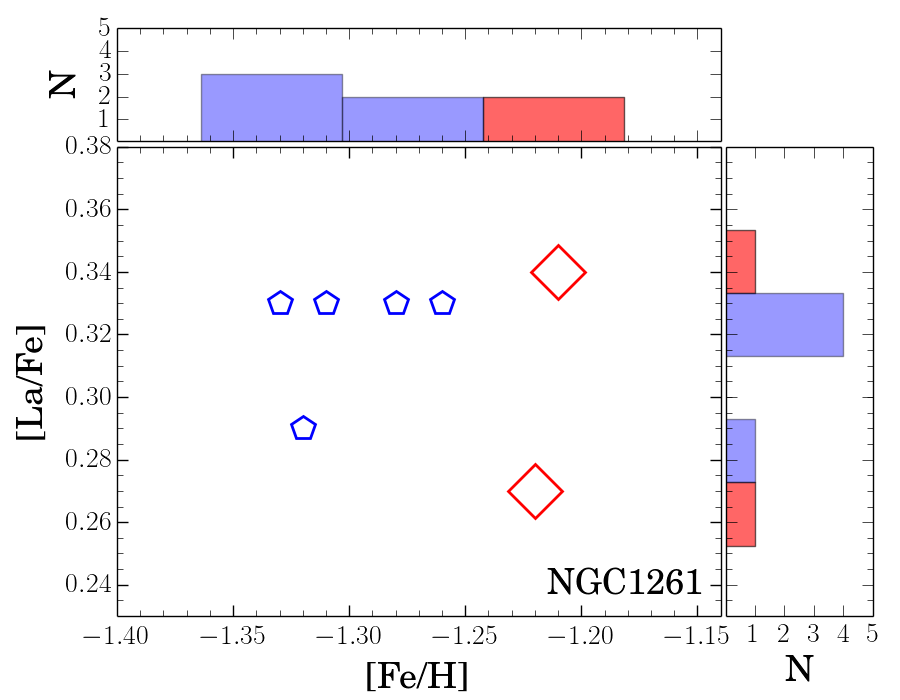}
     \caption{ Lanthanum abundance as a function of [Fe/H] for NGC\,6934 (left) and NGC\,1261 (right). Blue- and red-sequence stars are represented with blue and red symbols, respectively. The histograms represent the distribution of [Fe/H] and [La/Fe] of stars in the corresponding ChM sequence. }
       \label{fig:LaFe}
  \end{figure*}
\end{center}

Again, we do not find any evidence of internal variation in the $n$-capture
elements, as also suggested by the [La/Fe] versus [Fe/H] plot displayed in
Figure~\ref{fig:LaFe}. 

\begin{center}
  \begin{figure}
  \includegraphics[width=8.cm]{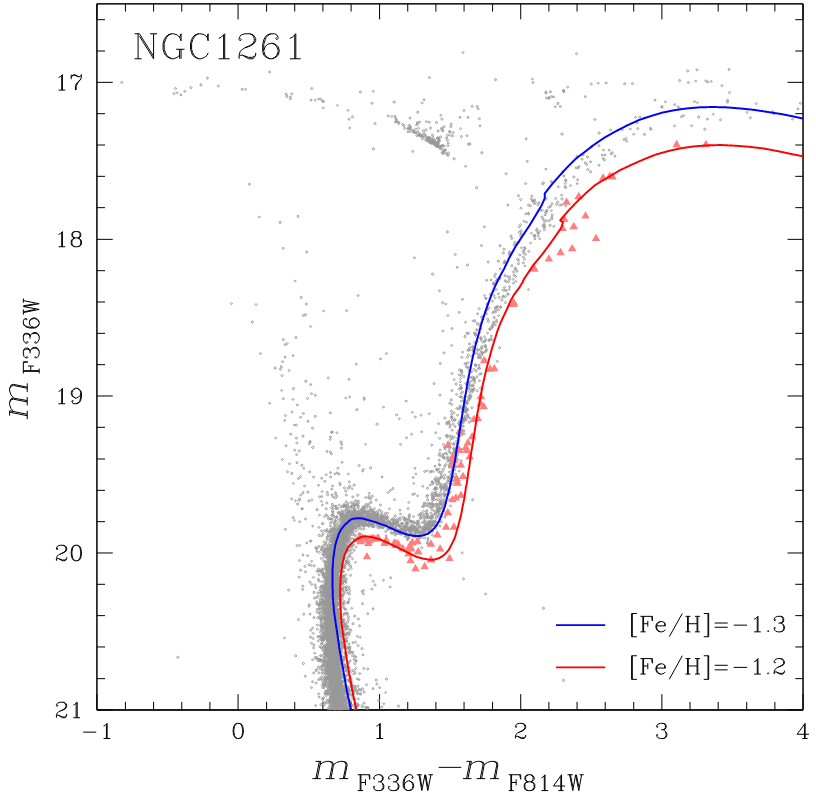}
     \caption{Comparison between the CMD of NGC\,1261 and isochrones from \citet{dotter2008a}. 
     Faint-SGB and red-RGB stars are represented with red symbols. 
     The red and blue isochrone have the same ages of 12.75 Gyr but different iron abundances of [Fe/H]=-1.2 and [Fe/H]=-1.3, respectively.} 
       \label{fig:ISO1261}
  \end{figure}
\end{center}

The homogeneous content of $s$-elements may provide further constraints on iron variations in NGC\,1261. Indeed, $s-$element enhancement is likely due to intermediate-mass AGB stars and is associated with  enhancement in the overall C$+$N$+$O content \citep[e.g.][]{ventura2009a}. The fact that blue-RGB and red-RGB stars share the same [La/Fe] content may suggest that all NGC\,1261 stars share the same overall CNO abundance.  
By assuming constant C$+$N$+$O, iron variation is possibly the only responsible for the SGB split. As shown in Figure \ref{fig:ISO1261}, the  faint SGB and the red RGB of NGC\,1261 are consistent with an  isochrone with the same age (12.75 Gyr) and $[\alpha/Fe]=0.4$ as the blue-RGB and the bright SGB, but enhanced in [Fe/H] by 0.1 dex. This result corroborates the finding from spectroscopy that red-RGB stars are more iron-rich than the blue RGB.

%

\section{Discussion and conclusions}\label{sec:conclusions}

The chemical enrichment in iron has been found to be a feature of
Type~II GCs, which have the stars distributed on the redder side of
the ChM enhanced in the 
 metallicity. The fact that Type~II GCs
are among the most massive clusters in the Milky Way is reminiscent of
the idea that more massive objects were more efficient in retaining
the ejecta from SNe, though not all massive GCs are Type\,II \citep[][]{Mil17}.

\begin{center}
  \begin{figure}
  \includegraphics[width=9cm]{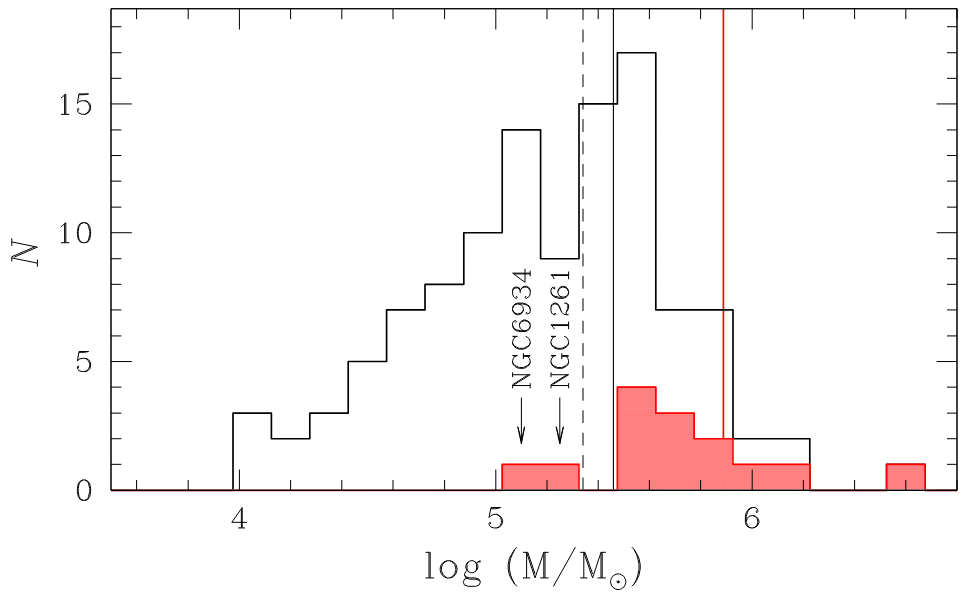}
     \caption{Histogram distributions of 
     present-day masses for Milky-Way GCs (black) and for Type II GCs (red-filled histogram). Gray-dashed and red-continuous vertical lines indicate the average masses 
      of these two groups of GCs, whereas the continuous vertical line mark the average mass of all clusters. GC masses are from \citet{baumgardt2018a}.}
       \label{fig:mass}
  \end{figure}
 \end{center}

Figure~\ref{fig:mass} shows the distributions in 
 present-day masses of Galactic GCs, with Type~II
GCs indicated by the red-filled histogram. 
One can see that NGC\,6934 and NGC\,1261 are 
identified so far, but still more massive than 10$^{5} M_{\odot}$. 
 
For NGC\,6934 we confirm that the stars on the red side of the map
have a iron abundances higher than the bulk of the stars in the
clusters defining the blue sequence on the ChM. The observed enrichment in Fe is
$\Delta$[Fe/H]$=+$0.21$\pm$0.04~dex. 
Some hint of a small Fe enrichment, by $\Delta$[Fe/H]$=+$0.08$\pm$0.01~dex
has been detected also in the red RGB of NGC\,1261, although having only two stars makes it difficult to draw strong conclusions from spectroscopy alone.

However, we notice that the analyzed stars are homogeneous in $s-$process elements, which would suggest that red- and blue-RGB stars have constant C$+$N$+$O content \citep[e.g.][]{cassisi2008a, ventura2009a, yong2009a}.  If this hypothesis is correct, the only way to reproduce the photometric splits on the SGB and the RGBs is to assume that red-RGB/faint-SGB stars are enhanced in [Fe/H] by ~0.1. This fact would corroborate the spectroscopic evidence of a metallicity variation in NGC\,1261. 

Following the approach introduced in \citet{renzini2008a}, we calculate for these two and eight more Type II GCs the excess iron content (F$_{\rm plus}$) of the red sequence population in these clusters. This is defined as:
\begin{equation}
    F_{\rm plus}=Z_{\rm Fe}^{\rm R}-Z_{\rm Fe}^{\rm B} M_{\rm R}
\end{equation}
where $Z_{\rm Fe}^{\rm R}$ and $Z_{\rm Fe}^{\rm B}$ are the iron abundances mass in the blue sequence and in the red sequence, respectively, and $M_{\rm R}$ is the mass of the red sequence. As suggested by  \citet{renzini2008a}, this quantity can be related to the iron produced by the core collapse supernovae of blue-sequence population. 

\begin{center}
  \begin{figure*}
  \includegraphics[width=15cm]{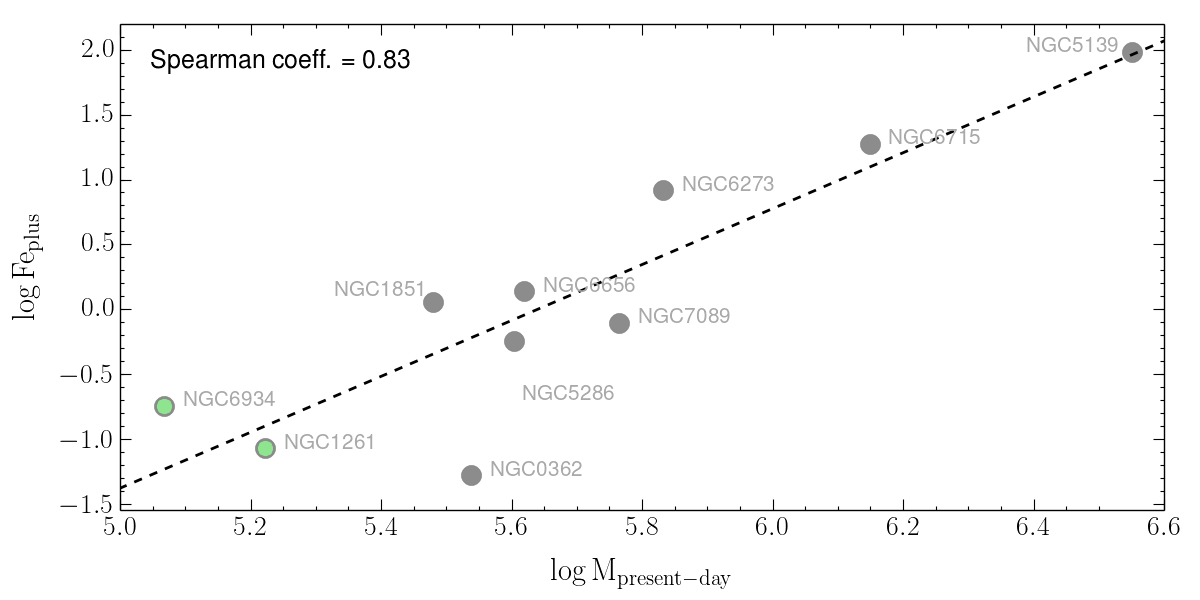}
  \includegraphics[width=15cm]{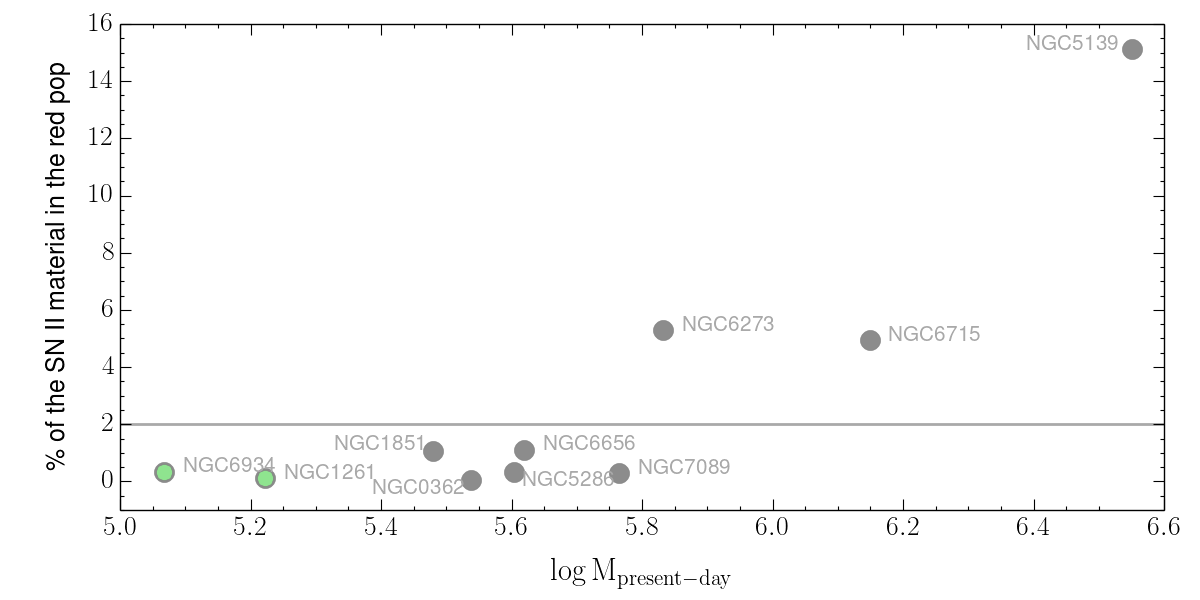}
     \caption{
     Logarithm of the additional Fe ($\rm {log~Fe_{plus}}$) in the
     metal richer stellar populations  as a function of the logarithm
     present-day GC masses. 
     $\rm {Fe_{plus}}$ has been calculated as the additional mass
     fraction of Fe multiplied by the fraction of mass in the redder
     stars \citep[see Table 2 from][]{Mil17}. 
     Fraction of iron produced by SNe~II that has been retained to form
     the populations enhanced in iron.
     }
       \label{figConclusions}
  \end{figure*}
 \end{center}

Figure~\ref{figConclusions} displays the additional Fe in the red
stars of ten Type~II GCs as a function of their present-day mass, with
both quantities on a logarithmic scale. As discussed above, the values $\rm {Fe_{plus}}$
have been calculated as the Fe mass fraction of the red stars on the
ChMs multiplied by the stellar mass in the corresponding
population\footnote{In the case of more than one ``red sequence'' on the
  ChM, we have considered all the stellar populations. Given the
  difficulty to select individual stellar populations in
  $\omega$~Centauri, for this GC we have considered the
  36\% of stars with [Fe/H]=$-$1.85, the 50\% with [Fe/H]=$-$1.50, and
  the 14\% with [Fe/H]=$-$1.00.}. 
There is a quite clear correlation among the two plotted quantities,
with the more massive GCs showing higher $\rm {Fe_{plus}}$, which
means a higher Fe-enriched stellar masses. NGC\,6934 and NGC\,1261,
together have the lowest Fe-enriched mass. A possible exception is NGC\,362, which is consistent with having constant iron abundance \citep{Mar19}. 

Translating the $\rm {Fe_{plus}}$ in each cluster into the fraction of Fe
produced by SNe~II that has been retained to form the Fe-richer
stars\footnote{This quantity has been derived assuming that a stellar
  generation makes $\sim$0.5~$\rm {M_{\odot}}$  of iron from core
  collapse supernovae every 1,000 $\rm {M_{\odot}}$ of gas turned into
  stars \citep[e.g.\,][and references therein]{Renzini&Andreon}.}, 
we find that only three clusters, namely NGC\,6273, NGC\,6715 (M\,54)
and NGC\,5139 ($\omega$~Centauri), have retained more than the 2\% of
the SNe~II material. Not surprisingly, $\omega$~Centauri is the GC
that has retained most material from Supernovae, NGC\,6273 and M\,54
have retained around the 5\% of metal-enriched SNe ejecta, NGC\,1851
and NGC\,6656 (M\,22) between 1 and 2\%. The clusters analysed
here, NGC\,6934 and NGC\,1261 are among those that retained a lower
amount of material.

We conclude that typically only a few percent of the iron produced by the blue sequence population was incorporated in the red sequence population, while all the rest was lost by the system. 

Interestingly enough, the two less massive Type~II GCs analysed so
far in the context of ChM stellar populations, namely NGC\,6934 and
NGC\,1261, are the only known Type~II
GCs with no evidence for variations in the $s$-element chemical
abundances. 
This finding may add new insights into the understanding of this class of globulars suggesting that a mass threshold could
exist either for the retention of the $s$-process enriched material
and/or for the capability of sustaining more-prolonged star formation
histories with contributions from lower-mass AGB stars.

We conclude by emphasising 
 that the total stellar mass of the cluster is indeed a  fundamental parameter for shaping the pattern of multiple stellar
populations in GCs \citep[see also][]{Mil17, milone2020a, lagioia2019a, dondoglio2021a}. This parameter is also relevant within the
GCs associated to the newly-discovered class of {\it anomalous} Type~II GCs:
\begin{itemize}

\item{Type~II GCs lie along the more massive side of the Milky-Way GC mass distribution.}

\item{Among these clusters, the more-massive the GC is, the higher is the amount of iron that has been incorporated in the red {\it anomalous} stellar population on the ChM.}

\item{Only some GCs with present-day total masses ${M}\gtrsim 10^{5.8} M_{\odot}$ have   retained a fraction of SNII material higher than the 2\%.}

\item{At odds with the majority of {\it anomalous}/Type~II GCs, the two less massive Type~II GCs analysed so far show no evidence of $s$-processed based enrichment.}
\end{itemize}

\section{Appendix. The Lanthanum abundances along the Chromosome Map of
  M\,54}\label{sec:M54} 

Variations in $s$-process elements have been observed in various Type\,II GCs, and are often associated with variations in the overall CNO content \citep[e.g.][]{Y&G08, yong2009a, Mar09}. Although M\,54 is one the most studied Type\,II GCs,  the abundance of $s-$elements in its stellar populations has been never investigated.  
In this appendix we use the Fe abundances and atmospheric parameters from \citet{car10b} to derive first abundances of Lanthanum by exploiting the GIRAFFE spectra analysed by the same authors.

\begin{figure*}
  \includegraphics[width=18.4cm]{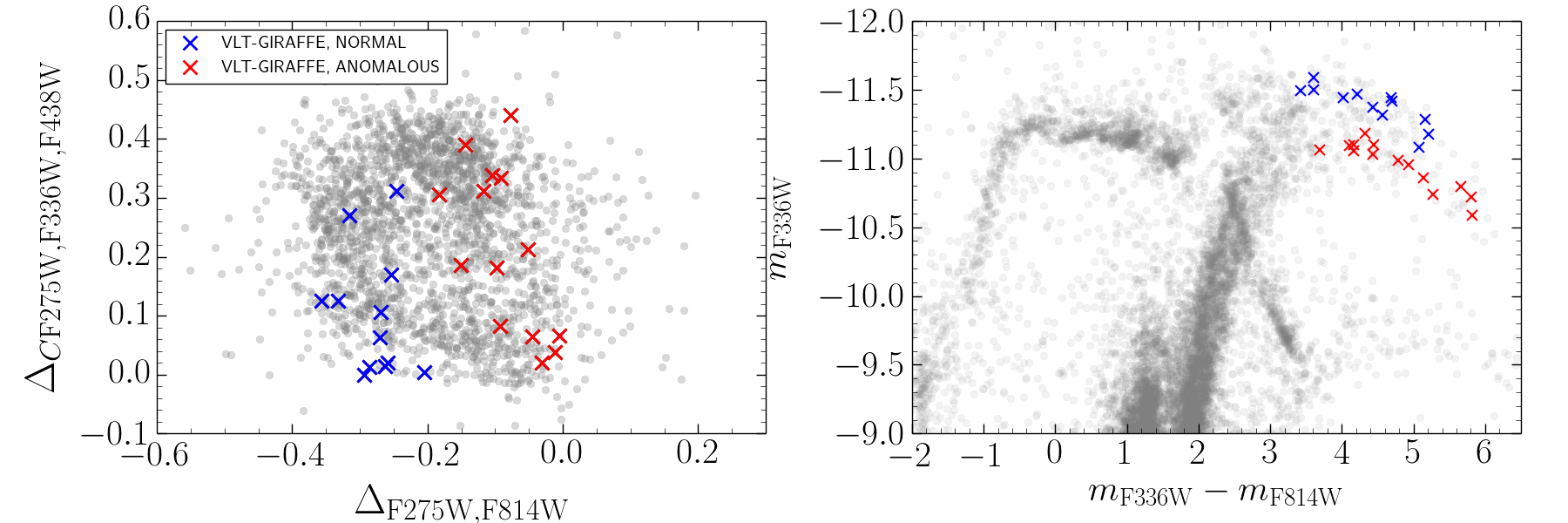}
     \caption{Chromosome map (left panel) and $m_{\rm F336W}$ vs.\, $m_{\rm
         F336W}-m_{\rm F814W}$ CMD (right panel) for M\,54. 
      The stars coloured in red have been selected as the stars defining the
      redder RGB stars on the CMD. Blue and red crosses are the
      GIRAFFE targets of \citet{car10b}, that we have associated with
      the red and blue RGB/ChM sequence, respectively.}
       \label{fig:PhotM54}
  \end{figure*}

\begin{figure}
  \includegraphics[width=8.65cm]{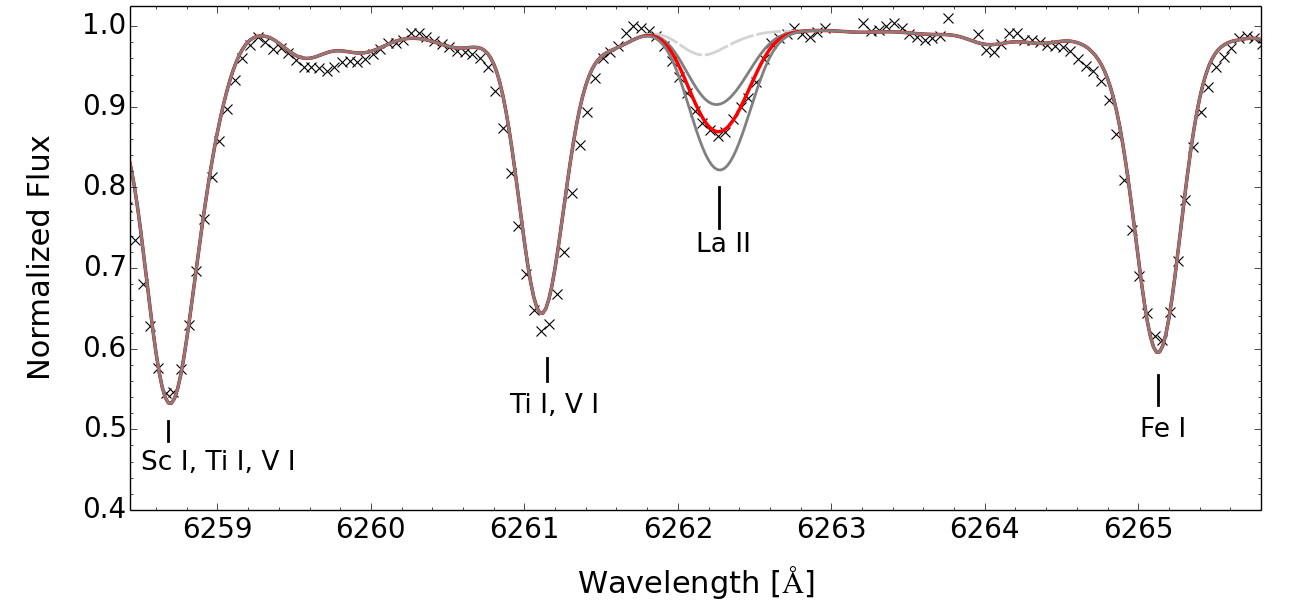}
     \caption{Observed and synthetic spectra around the La line at
       6262~\AA\ for star \#38000997 in M\,54 (GIRAFFE data from 
       \citet{car10b}). The cross points represent the observed spectrum,
       the light-grey dashed line is the spectrum computed with no
       contribution from La; the red line is the best-fitting
       synthesis; and the grey lines are the synthetic spectra computed with
       La abundances altered by  $\pm$0.2~dex from the best value. 
       For this synthesis we have assumed [N/Fe]=$+$0.8~dex.}
       \label{fig:SpettriM54}
  \end{figure}

Iron abundances, together with light elements, of M\,54 along the ChM have been already investigated in \citet{Mar19}, who identified eight stars on the blue RGB and ten stars on the red one. They found mean [Fe/H]$=-$1.73$\pm$0.06 (r.m.s.=0.16), and
[Fe/H]$=-$1.39$\pm$0.03 (r.m.s.=0.11).
Figure~\ref{fig:PhotM54} shows the ChM of M\,54 and the $m_{\rm
  F336W}$ vs.\,$m_{\rm F336W}-m_{\rm F814W}$ CMD for the stars observed
with GIRAFFE. Here we identify 12 stars on the blue RGB, and 14 on the
red one. The presence of more stars with available ChM data than in
\citet{Mar19} is basically due to our more generous photometric
constraints which include a larger range in magnitudes. 
The zoom in of a typical spectrum around the La line at 6262 \AA\, has been plotted in Figure \ref{fig:SpettriM54}, where we also show the synthetic spectra that provide the best fit with the observations, an the synthetic spectra that differ in [La/Fe] by $\pm$0.2 dex from the best-fit ones. 

To derive the internal error associated with the La measurements of M\,54, we
used the uncertainties in the atmospheric parameters of
\citet{car10b}, i.e.\,
$\Delta$\teff/$\Delta$\logg/$\Delta$[A/H]/$\Delta$\vmicro=\\
$\pm$50~K/$\pm$0.2~dex/$\pm$0.1~dex/$\pm$0.1\kmsec. \\
As expected, [La/Fe] abundances are mostly affected by the uncertainties in the surface gravities, as the values vary by
$\pm$0.08 by changing \logg\ by $\pm$0.20~dex. Variations in
metallicity by $\pm$0.10 change [La/Fe] by $\pm$0.05, while
temperature and microturbulence do not introduce significant
uncertainties to the results.
The quality of the data affects the results in [La/Fe] by 0.10~dex. 
The total estimated uncertainty associated with our individual La abundances is
0.14~dex, which is comparable with the r.m.s. obtained for the mean La
abundance of the blue and red-RGB stars in M\,54.

\begin{figure*}
  \includegraphics[width=18.4cm]{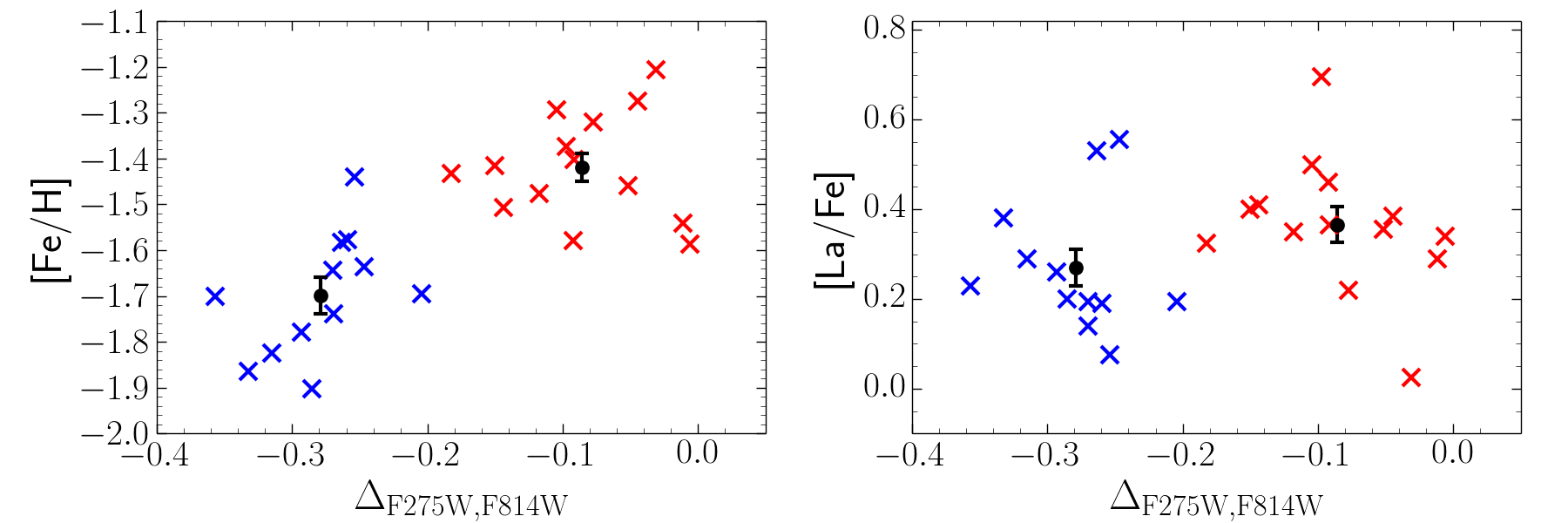}
     \caption{[Fe/H] and [La/Fe] abundances as a function of the \x\
       axis of the ChM for M\,54. Blue and red crosses represent the
       stars associated with the red and blue RGB/ChM sequence,
       respectively, as in Figure~\ref{fig:PhotM54}. In each panel,
       we indicate the mean abundances, with the asscociated error,
       for each group.  
      }
       \label{fig:FeM54}
  \end{figure*}

Results are shown in Figure~\ref{fig:FeM54}, where we represent the [Fe/H] and [La/Fe] as a function of
the \x\ values. As expected, the distribution of stars along \x\ is
sensitive to metallicity. Our mean [Fe/H] for the stars on the
ChM of M\,54 are [Fe/H]$=-$1.70$\pm$0.04 (r.m.s.=0.13) and
[Fe/H]$=-$1.42$\pm$0.03 (r.m.s.=0.12), for the blue and red RGB,
respectively.
Lanthanum over iron might be marginally enhanced among Fe-rich stars, being the
mean values for blue and red RGBs: [La/Fe]$=+$0.27$\pm$0.04
(r.m.s.=0.15) and [La/Fe]$=+$0.37$\pm$0.04 (r.m.s.=0.15),
respectively. However, with the current dataset the difference is only
at 1.5~$\sigma$ level, preventing any strong conclusion on [La/Fe] variations between red- and  blue-RGB stars.
On the other side, the absolute abundance of La in the two RGBs is significantly different and ranges from [La/H]=$-1.43$ in the blue RGB to [La/H]=$-1.05$ in the red RGB.

Lanthanum variations are common features of massive  Type\,II GCs, but the [La/Fe] difference between red-RGB and blue-RGB stars seems unrelated to the [Fe/H] difference. [La/Fe] varies by $\sim$0.2 dex \citep[in M\,2][]{Yong14} up to $\sim$1.5 dex \citep[in $\omega$\,Centauri,][]{Mar12_M22}. Lanthanum variations of $\sim$0.3 dex   are observed in both NGC\,1851, which exhibits tiny Fe differences between red- and blue-RGB stars ($\sim$0.05 dex) and NGC\,6656 where we observe a small iron spread relative to M\,54 of $\sim$0.15 dex \citep[][]{Y&G08, car10, Mar09, Mar11, dacosta2009a}.  Similarly, NGC\,5286 shows a large [La/Fe] difference of $\sim$0.65 dex, and iron variation of $\sim 0.2$ dex \citep[][]{Mar15}, whereas the large iron variation of NGC\,6273 \citep[$\sim$0.5 dex,][]{johnson2017a} corresponds to a [La/Fe] difference of $\sim$0.4 dex.  
On the other hand, red-RGB and blue-RGB stars of the Type II GC NGC\,362 have the same iron abundance, but different content of s-process elements \citep[][]{Mar19}.

The small [La/Fe] variation observed in M\,54, which is one of the studied Type\,II GC with the biggest iron variation, suggests that multiple processes govern the chemical enrichment in the Type II GCs. The s-process enrichment, possibly due to AGB stars,  seems to be poorly efficent in M\,54 despite the significant [Fe/H] spread, possibly driven by supernovae.  

                        
%
%
%
%
%

\acknowledgments
This work has received funding from the European Research Council (ERC) under the European Union's Horizon 2020 research innovation programme (Grant Agreement ERC-StG 2016, No 716082 'GALFOR', PI: Milone, http://progetti.dfa.unipd.it/GALFOR).
APM acknowledges support from MIUR through the FARE project R164RM93XW SEMPLICE (PI: Milone). APM has been supported by MIUR under PRIN program 2017Z2HSMF (PI: Bedin).


\begin{deluxetable}{cccccccccc}
\tablecaption{
Observation details, coordinates and radial velocities, with associated
r.m.s. and number of exposures, for our spectroscopic targets observed
with UVES and GIRAFFE.\label{tab:obs}}
\tablehead{
  GC       & STAR  &  RA & DEC  & mag & RV$_{\rm helio}$    & r.m.s. &\# exposures & SOURCE &Alternative ID\\
           &       &J2000& J2000&     &(\kmsec)            & (\kmsec)&            &        &\citep{Mar19}}
\startdata
NGC\,6934  &  37   & 20:34:12.28 & 07:24:16.3 &     & $-$409.74$\pm$0.05 & 0.19   & 9 & UVES & \\
NGC\,6934  &  29   & 20:34:14.37 & 07:24:49.3 &     & $-$414.88$\pm$0.03 & 0.14   & 9 & UVES & \\
NGC\,6934  &  27   & 20:34:16.46 & 07:24:53.3 &     & $-$402.72$\pm$0.06 & 0.23   & 9 & UVES & NGC6934-a1\\
NGC\,6934  &  14   & 20:34:13.66 & 07:24:07.2 &     & $-$414.85$\pm$0.14 & 0.57   & 9 & UVES & \\
NGC\,6934  &  1    & 20:34:13.80 & 07:22:47.8 &     & $-$408.36$\pm$0.05 & 0.22   & 9 & UVES & \\
NGC\,6934  &  5    & 20:34:10.34 & 07:23:21.2 &     & $-$412.25$\pm$0.08 & 0.35   & 9 & UVES & \\
NGC\,6934  &  47   & 20:34:07.58 & 07:24:17.1 &     & $-$402.50$\pm$0.06 & 0.26   & 9 & UVES & NGC6934-a2\\\hline
%
NGC\,6934  &  22   & 20:34:09.77 & 07:24:09.5 &     & $-$411.52$\pm$0.05 & 0.15   & 9 & GIRAFFE &  \\
NGC\,6934  &  23   & 20:34:09.72 & 07:23:27.5 &     & $-$407.54$\pm$0.04 & 0.13   & 9 & GIRAFFE &  \\
NGC\,6934  &  34   & 20:34:12.63 & 07:24:40.4 &     & $-$405.94$\pm$0.09 & 0.27   & 9 & GIRAFFE &  \\
NGC\,6934  &  35   & 20:34:12.63 & 07:24:57.1 &     & $-$407.24$\pm$0.05 & 0.15   & 9 & GIRAFFE &  \\
NGC\,6934  &  42   & 20:34:10.39 & 07:24:17.8 &     & $-$411.15$\pm$0.04 & 0.11   & 9 & GIRAFFE & NGC6934-n1 \\
NGC\,6934  &  50   & 20:34:11.89 & 07:25:26.7 &     & $-$403.68$\pm$0.12 & 0.35   & 9 & GIRAFFE &  \\\hline
%
NGC\,1261  &  14    &03:12:19.83 &$-$55:13:08.0 &     & 65.89$\pm$0.09 & 0.34   & 7 & UVES & \\
NGC\,1261  &  15    &03:12:19.60 &$-$55:13:42.6 &     & 72.61$\pm$0.41 & 1.49   & 7 & UVES & \\
NGC\,1261  &  17    &03:12:18.25 &$-$55:13:31.0 &     & 76.49$\pm$0.10 & 0.35   & 7 & UVES & \\
NGC\,1261  &  36    &03:12:21.40 &$-$55:12:24.5 &     & 75.62$\pm$0.12 & 0.42   & 7 & UVES & \\
NGC\,1261  &  49    &03:12:15.44 &$-$55:12:21.3 &     & 67.07$\pm$0.07 & 0.24   & 7 & UVES & \\
NGC\,1261  &  52    &03:12:09.78 &$-$55:12:49.9 &     & 65.28$\pm$0.06 & 0.22   & 7 & UVES & \\
NGC\,1261  &   6    &03:12:12.87 &$-$55:13:52.1 &     & 72.20$\pm$0.15 & 0.55   & 7 & UVES & \\\hline
NGC\,1261  &   7    &03:12:10.53 &$-$55:14:07.4 &     & 74.52$\pm$0.16 & 0.39  & 7 & GIRAFFE & \\
NGC\,1261  &   22   &03:12:14.88 &$-$55:13:40.4 &     & 74.38$\pm$0.10 & 0.24  & 7 & GIRAFFE & \\
NGC\,1261  &   24   &03:12:13.62 &$-$55:13:24.8 &     & 74.24$\pm$0.16 & 0.38  & 7 & GIRAFFE & \\
NGC\,1261  &   35   &03:12:25.58 &$-$55:12:53.3 &     & 69.79$\pm$0.10 & 0.25  & 7 & GIRAFFE & \\
NGC\,1261  &   40   &03:12:19.31 &$-$55:12:33.3 &     & 66.57$\pm$0.10 & 0.24  & 7 & GIRAFFE & \\
NGC\,1261  &   54   &03:12:22.86 &$-$55:12:04.4 &     & 69.82$\pm$0.09 & 0.23  & 7 & GIRAFFE & \\
NGC\,1261  &   61   &03:12:13.49 &$-$55:11:47.7 &     & 70.60$\pm$0.08 & 0.19  & 7 & GIRAFFE & \\\hline
\enddata
\end{deluxetable}
%

\begin{deluxetable}{cccccccccc}
\tablewidth{12pt}
\tablecaption{
Atmospheric parameters for our spectroscopic targets observed
with UVES and GIRAFFE.\label{tab:atm}}
\tablehead{
  GC       & STAR  & \teff & \logg & [Fe/H] & \vmicro & \teff$_{\rm phot}$ &\logg & \vmicro&SOURCE \\
           &       &       &       &        &         &  & &&        }
\startdata                         
NGC\,6934  &  37   &  4280 & 1.03  & $-$1.36& 1.55    & 4214 &0.90 &1.70 &UVES \\
NGC\,6934  &  29   &  4620 & 1.72  & $-$1.34& 1.45    & 4580 &1.59 &1.52 &UVES \\
NGC\,6934  &  27   &  4490 & 1.50  & $-$1.30& 1.51    & 4453 &1.41 &1.57 &UVES \\
NGC\,6934  &  14   &  4350 & 0.70  & $-$1.60& 2.05    & 4279 &0.87 &1.71 &UVES \\
NGC\,6934  &  1    &  4240 & 0.90  & $-$1.53& 1.90    & 4165 &0.72 &1.75 &UVES \\
NGC\,6934  &  5    &  4050 & 0.43  & $-$1.52& 2.20    & 4029 &0.42 &1.82 &UVES \\
NGC\,6934  &  47   &  4430 & 1.23  & $-$1.35& 1.63    & 4379 &1.22 &1.62 &UVES \\\hline
NGC\,6934   &  22   &  4545 & 1.46  &$-$1.62 &  1.22  & 4545 &1.46 & 1.56 & GIRAFFE  \\
NGC\,6934   &  23   &  4485 & 1.37  &$-$1.64 &  1.50  & 4485 &1.37 & 1.58 & GIRAFFE  \\
NGC\,6934   &  34   &  4644 & 1.71  &$-$1.69 &  1.33  & 4644 &1.71 & 1.50 & GIRAFFE  \\
NGC\,6934   &  35   &  4432 & 1.20  &$-$1.56 &  1.60  & 4432 &1.20 & 1.62 & GIRAFFE  \\
NGC\,6934   &  42   &  4342 & 1.05  &$-$1.58 &  1.67  & 4342 &1.05 & 1.66 & GIRAFFE  \\
NGC\,6934   &  50   &  4735 & 1.92  &$-$1.61 &  1.08  & 4735 &1.92 & 1.44 & GIRAFFE  \\
%
\hline
NGC\,1261  &  14   & 4400  &1.15   & $-$1.26&  1.68   & 4337 &1.13 & 1.64& UVES \\
NGC\,1261  &  15   & 4150  &0.70   & $-$1.31&  1.90   & 4107 &0.71 & 1.75& UVES \\
NGC\,1261  &  17   & 4470  &1.40   & $-$1.22&  1.62   & 4469 &1.38 & 1.58& UVES \\
NGC\,1261  &  36   & 4300  &1.05   & $-$1.28&  1.80   & 4255 &0.98 & 1.68& UVES \\
NGC\,1261  &  49   & 4120  &0.60   & $-$1.33&  1.95   & 4106 &0.69 & 1.76& UVES \\
NGC\,1261  &  52   & 4570  &1.60   & $-$1.21&  1.61   & 4521 &1.49 & 1.55& UVES \\
NGC\,1261  &   6   & 4130  &0.60   & $-$1.32&  1.93   & 4094 &0.68 & 1.76& UVES \\\hline
NGC\,1261  &  7    & 4738  & 2.03  & $-$1.38& 1.32    & 4738 &2.03 &1.41 &GIRAFFE \\
NGC\,1261  &  22   & 4544  & 1.59  & $-$1.38& 1.59    & 4544 &1.59 &1.53 &GIRAFFE \\
NGC\,1261  &  24   & 4185  & 0.89  & $-$1.31& 1.82    & 4185 &0.89 &1.70 &GIRAFFE \\
NGC\,1261  &  35   & 4691  & 1.92  & $-$1.41& 1.32    & 4691 &1.92 &1.44 &GIRAFFE \\
NGC\,1261  &  40   & 4411  & 1.35  & $-$1.30& 1.64    & 4411 &1.35 &1.59 &GIRAFFE \\
NGC\,1261  &  54   & 4720  & 2.00  & $-$1.32& 1.17    & 4720 &2.00 &1.42 &GIRAFFE \\
NGC\,1261  &  61   & 4460  & 1.42  & $-$1.32& 1.53    & 4460 &1.42 &1.57 &GIRAFFE \\
\enddata
\end{deluxetable}
%
	
\newpage
\movetabledown=5.5cm
\begin{rotatetable}
\floattable
\begin{deluxetable}{cccccccccccccccccccccccc}
\tablewidth{6pt}
\tabletypesize{\scriptsize}
\tablecaption{Analyzed Chemical Abundances from O to Sc.\label{tab:LiToSc}}
\tablehead{
STAR &[O/Fe]&$\sigma$&\#&[Na/Fe]&$\sigma$&[Na/Fe] &$\sigma$ &\# & [Mg/Fe] & $\sigma$ & \# & [Al/Fe] & $\sigma$ & \# & [Si/Fe] & $\sigma$ & \# & [Ca/Fe] & $\sigma$ & \# & [Sc/Fe]{\sc ii} & $\sigma$ & \# \\
     &      &        &  &LTE    &        &non-LTE &         &   &         &          &    &         &          &    &         &          &    &         &          &    &         &          &          
}
\startdata
NGC\,6934-37   & 0.55 & 0.05 & 2 &    0.18 & 0.04 &   0.10  &0.01     & 4 & 0.33 & 0.03 & 2 & $-$0.10 & 0.11 & 2 & 0.36 & 0.10 & 8 & 0.32 & 0.14 & 20 & 0.02 & 0.14 & 7 \\
NGC\,6934-29   & 0.57 & 0.17 & 2 &    0.53 & 0.11 &   0.41  &0.08     & 4 & 0.33 & 0.05 & 2 &      -- &   -- & - & 0.35 & 0.06 & 5 & 0.29 & 0.12 & 20 & 0.07 & 0.19 & 6 \\
NGC\,6934-27   & 0.70 & 0.13 & 2 & $-$0.23 & 0.09 &$-$0.30  &0.07     & 4 & 0.32 & 0.04 & 2 &    0.00 &   -- & 1 & 0.36 & 0.10 & 5 & 0.33 & 0.13 & 20 & 0.00 & 0.14 & 7 \\
NGC\,6934-14   & 0.44 & 0.03 & 2 &    0.32 & 0.02 &   0.24  &0.01     & 4 & 0.43 & 0.12 & 2 &    0.29 & 0.01 & 2 & 0.42 & 0.09 & 8 & 0.25 & 0.10 & 20 & $-$0.03 & 0.11 & 7 \\
NGC\,6934-1    & 0.51 & 0.01 & 2 &    0.26 & 0.03 &   0.18  &0.02     & 4 & 0.43 & 0.04 & 2 &    0.20 & 0.02 & 2 & 0.39 & 0.11 & 7 & 0.28 & 0.07 & 20 & 0.02 & 0.11 & 7 \\
NGC\,6934-5    & 0.62 & 0.06 & 2 & $-$0.07 & 0.05 &$-$0.13  &0.03     & 4 & 0.45 & 0.07 & 2 & $-$0.07 & 0.02 & 2 & 0.44 & 0.11 & 7 & 0.23 & 0.09 & 20 & $-$0.04 & 0.10 & 7 \\
NGC\,6934-47   & 0.61 & 0.08 & 2 & $-$0.14 & 0.04 &$-$0.20  &0.03     & 4 & 0.40 & 0.08 & 2 & $-$0.17 &   -- & 1 & 0.34 & 0.08 & 7 & 0.33 & 0.11 & 20 & 0.04 & 0.12 & 7 \\\hline
avg. & 0.57 &      &   &    0.12 &      &   0.04  &         &   & 0.38 &      &   &    0.03 &      &   & 0.38 &      &   & 0.29  &      &    & 0.01 &      & \\
$\pm$& 0.03 &      &   &    0.11 &      &   0.11  &         &   & 0.02 &      &   &    0.08 &      &   & 0.02 &      &   & 0.02  &      &    & 0.02 &      & \\
$\sigma$&0.08&     &   &    0.28 &      &   0.26  &         &   & 0.06 &      &   &    0.18 &      &   & 0.04 &      &   & 0.04  &      &    & 0.04 &      & \\\hline
NGC\,1261-14   & 0.05 &  --  & 1 &    0.62 &  0.10 & 0.51  & 0.03    & 4 & 0.36  & 0.00  & 2 & 0.33 & 0.06  & 2  & 0.35  & 0.08  & 8 & 0.34  & 0.12  & 20  & 0.08  & 0.13  & 6  \\
NGC\,1261-15   & 0.39 & 0.04 & 2 &    0.49 &  0.10 & 0.38  & 0.04    & 4 & 0.42  & 0.06  & 2 & 0.21 & 0.03  & 2  & 0.43  & 0.11  & 6 & 0.28  & 0.10  & 19  & 0.08  & 0.11  & 6  \\
NGC\,1261-17   & 0.20 &  --  & 1 &    0.69 &  0.10 & 0.56  & 0.02    & 4 & 0.43  & 0.08  & 2 & 0.25 & 0.05  & 2  & 0.42  & 0.14  & 6 & 0.34  & 0.11  & 20  & 0.11  & 0.11  & 6  \\
NGC\,1261-36   & 0.48 & 0.07 & 2 &    0.37 &  0.11 & 0.27  & 0.05    & 4 & 0.41  & 0.02  & 2 & 0.13 & 0.06  & 2  & 0.38  & 0.08  & 5 & 0.29  & 0.10  & 20  & 0.06  & 0.12  & 6  \\
NGC\,1261-49   & 0.56 & 0.02 & 2 & $-$0.05 &  0.07 &$-$0.12& 0.04    & 4 & 0.45  & 0.03  & 2 &$-$0.09&0.04  & 2  & 0.44  & 0.11  & 6 & 0.26  & 0.10  & 19  & 0.04  & 0.11  & 5  \\
NGC\,1261-52   & 0.12 &  --  & 1 &    0.73 &  0.11 & 0.59  & 0.03    & 4 & 0.39  & 0.01  & 2 & 0.16 & 0.03  & 2  & 0.39  & 0.09  & 8 & 0.36  & 0.12  & 20  & 0.03  & 0.15  & 5  \\
NGC\,1261-6    & 0.55 & 0.02   & 2 & $-$0.06 &  0.07 &$-$0.13& 0.06  & 4 & 0.40  & 0.09  & 2 &$-$0.15& 0.03 &2   & 0.41  & 0.12  & 6 & 0.26  & 0.10  & 20  & 0.07  & 0.10  & 5  \\\hline
avg. & 0.34 &    &   &    0.40 &       & 0.29 &     &   & 0.41  &      &   & 0.12     &      &   & 0.40  &      &   & 0.30     &      &    & 0.07  &      & \\
$\pm$& 0.09 &    &   &    0.14 &       & 0.13 &     &   & 0.01  &      &   & 0.07     &      &   & 0.01  &      &   & 0.02     &      &    & 0.01  &      & \\
$\sigma$&0.21&   &   &    0.33 &       & 0.31 &     &   & 0.03  &      &   & 0.18     &      &   & 0.03  &      &   & 0.04     &      &    & 0.03  &      & \\\hline
\enddata
\end{deluxetable}
\end{rotatetable}

\newpage

\movetabledown=5.5cm
\begin{rotatetable}
\floattable
\begin{deluxetable}{cccccccccccccccccccccccccc}
\tablewidth{6pt}
\tabletypesize{\scriptsize}
\tablecaption{Analyzed Chemical Abundances from Ti to Ni.\label{tab:TiToNi}}
\tablehead{
STAR & [Ti/Fe]{\sc i} & $\sigma$ & \# & [Ti/Fe]{\sc ii} & $\sigma$ & \# & [V/Fe] & $\sigma$ & \# & [Cr/Fe]{\sc i} & $\sigma$ & \# & [Cr/Fe]{\sc ii} & $\sigma$ & \# & [Mn/Fe] & $\sigma$ & \# & [Co/Fe] & $\sigma$ & \# & [Ni/Fe] & $\sigma$ & \# }
\startdata
NGC\,6934-37 &0.21 &0.12 & 23 & 0.21 &0.04  & 5 &    0.08 & 0.09 & 14 &$-$0.17 &0.11 &5 &0.14 &0.04  & 2 & $-$0.52 & 0.11 & 5 &$-$0.07  & 0.08 & 2 &$-$0.02 &0.12 & 25 \\
NGC\,6934-29 &0.20 &0.14 & 19 & 0.21 &0.16  & 5 & $-$0.02 & 0.11 & 14 &$-$0.15 &0.05 &4 &0.19 & --   & 1 & $-$0.52 & 0.08 & 5 &$-$0.07  &   -- & 1 &$-$0.02 &0.10 & 21 \\
NGC\,6934-27 &0.18 &0.11 & 23 & 0.26 &0.13  & 5 & $-$0.03 & 0.09 & 14 &$-$0.13 &0.08 &4 &0.22 &0.07  & 2 & $-$0.51 & 0.07 & 5 &$-$0.09  & 0.07 & 2 &$-$0.04 &0.12 & 23 \\
NGC\,6934-14 &0.17 &0.09 & 22 & 0.26 &0.10  & 5 & $-$0.04 & 0.11 & 15 &$-$0.14 &0.11 &5 &0.01 &0.07  & 2 & $-$0.55 & 0.12 & 5 &$-$0.15  & 0.04 & 2 &$-$0.03 &0.09 & 25 \\
NGC\,6934-1  &0.22 &0.10 & 23 & 0.30 &0.06  & 5 &    0.05 & 0.08 & 14 &$-$0.15 &0.04 &5 &0.17 &0.06  & 2 & $-$0.47 & 0.08 & 5 &$-$0.07  & 0.03 & 2 &$-$0.03 &0.11 & 24 \\
NGC\,6934-5  &0.31 &0.14 & 22 & 0.22 &0.04  & 4 &    0.09 & 0.12 & 14 &$-$0.15 &0.09 &4 &0.16 &0.02  & 2 & $-$0.47 & 0.12 & 5 &$-$0.04  & 0.08 & 2 &   0.01 &0.11 & 25 \\
NGC\,6934-47 &0.20 &0.10 & 23 & 0.18 &0.07  & 5 &    0.01 & 0.08 & 14 &$-$0.08 &0.10 &4 &0.25 &0.07  & 2 & $-$0.47 & 0.07 & 5 &$-$0.05  & 0.01 & 2 &$-$0.03 &0.10 & 24 \\\hline
avg.&0.21&     &    & 0.23 &      &   &    0.02 &      &    &$-$0.14 &     &  &0.16 &      &   & $-$0.50 &      &   &$-$0.08  &      &   &$-$0.02 &     &    \\
$\pm$&0.02&    &    & 0.02 &      &   &    0.02 &      &    &   0.01 &     &  &0.03 &      &   &    0.01 &      &   &   0.01  &      &   &   0.01 &     &  \\
$\sigma$&0.05& &    & 0.04 &      &   &    0.05 &      &    &   0.03 &     &  &0.08 &      &   &    0.03 &      &   &   0.04  &      &   &   0.02 &     &  \\\hline
NGC\,1261-14 &0.25  &0.12 & 24 & 0.24 & 0.04 & 5 &  0.13& 0.11 & 14 &$-$0.14 & 0.08& 4&0.08 &0.08  & 2 &$-$0.41 &0.07  & 5 &$-$0.09  &0.06  & 2 &$-$0.03 &0.11 & 25 \\
NGC\,1261-15 &0.32  &0.11 & 21 & 0.27 & 0.08 & 4 &  0.28& 0.23 & 13 &$-$0.10 & 0.05& 4&0.16 &0.13  & 2 &$-$0.37 &0.09  & 5 &$-$0.05  &0.14  & 2 &$-$0.03 &0.11 & 24 \\
NGC\,1261-17 &0.25  &0.11 & 22 & 0.20 & 0.07 & 5 &  0.12& 0.10 & 13 &$-$0.11 & 0.06& 4&0.07 &0.05  & 2 &$-$0.44 &0.05  & 5 &$-$0.10  &0.05  & 2 &$-$0.04 &0.12 & 22 \\
NGC\,1261-36 &0.28  &0.12 & 24 & 0.24 & 0.06 & 5 &  0.19& 0.13 & 13 &$-$0.12 & 0.04& 4&0.13 &0.02  & 2 &$-$0.38 &0.07  & 5 &$-$0.01  &0.10  & 2 &   0.00 &0.10 & 25 \\
NGC\,1261-49 &0.32  &0.14 & 22 & 0.27 & 0.09 & 5 &  0.23& 0.22 & 13 &$-$0.13 & 0.09& 4&0.17 &0.08  & 2 &$-$0.42 &0.09  & 5 &$-$0.03  &0.08  & 2 &$-$0.02 &0.10 & 24 \\
NGC\,1261-52 &0.26  &0.12 & 23 & 0.25 & 0.07 & 5 &  0.09& 0.10 & 14 &$-$0.09 & 0.05& 3&0.15 &0.06  & 2 &$-$0.42 &0.04  & 5 &$-$0.13  &0.05  & 2 &   0.00 &0.11 & 22 \\
NGC\,1261-6  &0.33  &0.14 & 23 & 0.24 & 0.05 & 5 &  0.21& 0.23 & 14 &$-$0.12 & 0.09& 4&0.15 &0.14  & 2 &$-$0.40 &0.08  & 5 &$-$0.08  &0.15  & 2 &$-$0.03 &0.11 & 25 \\\hline
avg.&0.29&     &    & 0.24 &      &   &0.18 &      &    &$-$0.12 &     &  &0.13 &      &   &$-$0.41  &      &   & $-$0.07 &      &   &$-$0.02 &     &    \\
$\pm$&0.01&    &    & 0.01 &      &   &0.03 &      &    &0.01    &     &  &0.02 &      &   &  0.01   &      &   & 0.02    &      &   &   0.01 &     &  \\
$\sigma$&0.04& &    & 0.02 &      &   &0.07 &      &    &0.02    &     &  &0.04 &      &   &  0.02   &      &   & 0.04    &      &   &   0.02 &     &  \\
\enddata
\end{deluxetable}
\end{rotatetable}
\newpage
\movetabledown=5.5cm
\begin{rotatetable}
\floattable
\begin{deluxetable}{cccccccccccccccccccccccccccccccccc}
\tablewidth{6pt}
\tabletypesize{\scriptsize}
\tablecaption{Analyzed Chemical Abundances from Cu to Eu.\label{tab:CuToEu}}
\tablehead{
STAR& [Cu/Fe]&$\sigma$&\# & [Zn/Fe]& [Y/Fe] &$\sigma$&\#& [Zr/Fe] &[Ba/Fe]&$\sigma$&\# &[La/Fe]& $\sigma$&\# &[Ce/Fe]&$\sigma$&\#&[Pr/Fe]&[Nd/Fe]&$\sigma$&\#&[Eu/Fe]&$\sigma$&\# }
\startdata                          
NGC\,6934-37  & $-$0.64 & 0.08& 2& --   & $-$0.19 & 0.19 & 3 & 0.36 & 0.08 & 0.05 & 3 &  0.21  & 0.08 & 6 &0.10 &0.01&2 & 0.28 & 0.35 & 0.02 & 2 &0.62 &0.06 &2   \\
NGC\,6934-29  & $-$0.57 & 0.25& 2& --   & $-$0.12 & 0.25 & 3 & 0.25 & 0.22 & 0.05 & 3 &  0.29  & 0.09 & 5 & --  &--  &- & 0.39 & 0.20 & 0.01 & 2 &0.70 &--   &1   \\
NGC\,6934-27  & $-$0.56 & 0.16& 2& --   & $-$0.20 & 0.25 & 3 & 0.23 & 0.19 & 0.01 & 3 &  0.24  & 0.06 & 7 &0.13 &0.26&2 & 0.40 & 0.37 & 0.03 & 2 &0.61 &0.01 &2   \\
NGC\,6934-14  & $-$0.65 & 0.27& 2& --   & $-$0.11 & 0.07 & 3 & 0.20 & 0.19 & 0.04 & 3 &  0.19  & 0.09 & 7 &0.09 &0.17&2 & 0.23 & 0.22 & 0.08 & 2 &0.45 &0.08 &2   \\
NGC\,6934-1   & $-$0.46 & 0.12& 2& 0.03 & $-$0.11 & 0.14 & 3 & 0.40 & 0.25 & 0.04 & 3 &  0.28  & 0.08 & 7 &0.13 &0.05&2 & 0.34 & 0.33 & 0.00 & 2 &0.61 &0.02 &2   \\
NGC\,6934-5   & $-$0.51 & 0.08& 2& 0.14 & $-$0.09 & 0.22 & 3 & 0.40 & 0.18 & 0.02 & 3 &  0.26  & 0.06 & 7 &0.12 &0.02&2 & 0.35 & 0.45 & --   & 1 &0.62 &0.05 &2   \\
NGC\,6934-47  & $-$0.52 & 0.33& 2& --   & $-$0.13 & 0.16 & 3 & 0.31 & 0.17 & 0.03 & 3 &  0.25  & 0.08 & 7 &0.18 &--  &1 & 0.29 & 0.39 & 0.01 & 2 &0.59 &0.03 &2   \\ \hline
avg.& $-$0.56 & &&0.09 & $-$0.14 &      &   & 0.31 & 0.18 &      &   &  0.25  &      &   &0.13 &    &  & 0.33 & 0.33 &      &   &0.60 &     &  \\
$\pm$&   0.03 & &&0.11 &    0.02 &      &   & 0.03 & 0.02 &      &   &  0.01  &      &   &0.01 &    &  & 0.03 & 0.04 &      &   &0.03 &     &  \\
$\sigma$&0.07 & &&--   &    0.04 &      &   & 0.08 & 0.05 &      &   &  0.04  &      &   &0.03 &    &  & 0.06 & 0.09 &      &   &0.07 &     &  \\\hline
NGC\,1261-14  &$-$0.53& 0.39& 2 &$-$0.01& $-$0.09 & 0.14 & 3 & 0.23 & 0.17   & 0.03  & 3 & 0.33   & 0.13 &7  & 0.15 & 0.05& 2 & 0.35  & 0.43  & 0.09  & 2  & 0.63 & 0.00 & 2  \\
NGC\,1261-15  &$-$0.48& 0.22& 2 &   0.14& $-$0.06 & 0.26 & 3 & 0.26 &$-$0.07 & 0.06  & 3 & 0.33   & 0.10 &7  & 0.12 & 0.02& 2 & 0.42  & 0.55  &   --  & 1  & 0.66 & 0.04 & 2  \\
NGC\,1261-17  &$-$0.44& 0.26& 2 &   0.05& $-$0.10 & 0.17 & 3 & 0.22 & 0.15   & 0.03  & 3 & 0.27   & 0.13 &7  & 0.17 & 0.04& 2 & 0.36  & 0.43  & 0.02  & 2  & 0.63 & 0.02 & 2  \\
NGC\,1261-36  &$-$0.47& 0.23& 2 &   0.00& $-$0.12 & 0.18 & 3 & 0.27 & 0.09   & 0.07  & 3 & 0.33   & 0.12 &7  & 0.16 & 0.05& 2 & 0.43  & 0.54  & 0.02  & 2  & 0.70 & 0.03 & 2  \\
NGC\,1261-49  &$-$0.52& 0.16& 2 &   0.05& $-$0.03 & 0.22 & 3 & 0.30 &$-$0.06 & 0.08  & 3 & 0.33   & 0.09 &7  & 0.13 & 0.01& 2 & 0.40  & 0.54  &   --  & 1  & 0.69 & 0.03 & 2  \\
NGC\,1261-52  &$-$0.40& 0.21& 2 &    -- & $-$0.04 & 0.17 & 3 & 0.24 & 0.22   & 0.07  & 3 & 0.34   & 0.11 &7  & 0.17 & 0.01& 2 & 0.34  & 0.37  & 0.00  & 2  & 0.66 & 0.02 & 2  \\
NGC\,1261-6   &$-$0.45& 0.22& 2 &$-$0.02& $-$0.05 & 0.26 & 3 & 0.25 &$-$0.08 & 0.08  & 3 & 0.29   & 0.10 &7  & 0.12 & 0.00& 2 & 0.33  & 0.48  & 0.05  & 2  & 0.64 & 0.02 & 2  \\ \hline
avg.& $-$0.47 &&& 0.04 & $-$0.07 &      &   & 0.25 & 0.06 &     &   & 0.32  &      &   & 0.15&    &  & 0.38 & 0.48 &      &   & 0.66&     &  \\
$\pm$&0.02    &&& 0.03 &    0.01 &      &   & 0.01 & 0.05 &     &   & 0.01  &      &   & 0.01&    &  & 0.02 & 0.03 &      &   & 0.01&     &  \\
$\sigma$&0.05 &&& 0.06 &    0.03 &      &   & 0.03 & 0.13 &     &   & 0.03  &      &   & 0.02&    &  & 0.04 & 0.07 &      &   & 0.03&     &  \\
\enddata
\end{deluxetable}
\end{rotatetable}
\clearpage

\begin{deluxetable}{cc ccc ccc cc}
\tablewidth{12pt}
\tablecaption{
Chemical abundances obtained with GIRAFFE.\label{tab:abuGIRAFFE}}
\tablehead{
  GC       & STAR  & [O/Fe]&$\sigma$&\# & [Na/Fe]$_{\rm LTE}$& $\sigma$ & [Na/Fe]$_{\rm NLTE}$& $\sigma$ &\#    }
\startdata                         
NGC\,6934  &  22   &    -- & --     & - &  0.46  &   0.03   & 0.39   & 0.02& 2 \\ 
NGC\,6934  &  23   &   0.56&0.05    & 2 &  --    &    --    & --     &  -- & - \\ 
NGC\,6934  &  34   &   0.35& --     & 1 &  0.34  &   0.17   & 0.27   & 0.17& 2 \\ 
NGC\,6934  &  35   &   0.15& --     & 1 &  0.43  &   0.02   & 0.37   & 0.02& 2 \\ 
NGC\,6934  &  42   &   0.55&0.10    & 2 &  0.10  &   0.01   & 0.05   & 0.01& 2 \\ 
NGC\,6934  &  50   &   0.28& --     & 1 &  0.39  &   0.09   & 0.31   & 0.09& 2 \\ \hline
NGC\,1261  &   7   &   0.47&0.25    & 2 &$-$0.01 &   0.01   &$-$0.08 &0.01 & 2 \\
NGC\,1261  &  22   &   0.64&0.10    & 2 &  0.06  &   0.03   & 0.00   & 0.03& 2 \\
NGC\,1261  &  24   &   0.26&0.05    & 2 &  0.14  &   0.07   & 0.10   & 0.07& 2 \\
NGC\,1261  &  35   &   0.62&0.12    & 2 &$-$0.05 &   0.03   &$-$0.12 & 0.03 & 2 \\
NGC\,1261  &  40   &   0.31&0.11    & 2 &  0.26  &   0.04   & 0.21   & 0.04& 2 \\
NGC\,1261  &  54   &   --  & --     & 0 &  0.55  &   0.11   & 0.47   & 0.11& 2 \\
NGC\,1261  &  61   &   0.58&0.08    & 2 &$-$0.26 &   0.12   &$-$0.31 & 0.12& 2 \\
\enddata
\end{deluxetable}

\clearpage

\begin{table*}
\caption{This Table provides, for each Type\,II GC the average iron abundance of blue-RGB, red-RGB and extreme stars and the corresponding fractions of stars. We also provide the present-day GC mass \citep[from][]{baumgardt2018a}, the fraction of the iron produce by SNe that has been retained by the cluster, and the amount of iron produced by SNe. The last column provides the reference to the paper where iron abundances  of the normal and anomalous stars are derived. In NGC\,5286 only abundances from GIRAFFE data are used.}
\begin{tabular}{c cr cr cr r r ll}\hline
GC&[Fe/H]$\rm {_{blue\,RGB}}$&\%&[Fe/H]$\rm {_{red\,RGB}}$&\%&[Fe/H]$\rm {_{extreme}}$&\%&Mass &    SN	&Fe$_{\rm plus}$  &Reference \\
  & dex                   &   & dex                   &  & dex                    &  &[$\times 10^{5}\rm{M_{\odot}}$]&\%& M$_{\odot}$&\\\hline
NGC0362 &  $-$1.18 & 92.5 & $-$1.17 & 7.5 &   --    &  -- &  3.45   & 0.03  & 0.05207  &   \citet{Mar19}\\
NGC1261 &  $-$1.30 & 96.2 & $-$1.22 & 3.8 &   --    &  -- &  1.67   & 0.10  & 0.08415  &  This work	  \\
NGC1851	&  $-$1.19 & 70   & $-$1.13 & 30  &   --    &  -- &  3.02   & 1.07  & 1.133    &   \citet{car10}\\
NGC5286	&  $-$1.77 & 83.3 & $-$1.63 & 16.7&   --    &  -- &  4.01   & 0.34  & 0.5663   & \citet{Mar15} \\
NGC6273 &  $-$1.77 & 46   & $-$1.51 & 48  & $-$1.22 &   6 &  6.80   & 5.28  &  8.259   & \citet{johnson2017a}\\
NGC6656	&  $-$1.82 & 59.7 & $-$1.67 & 40.3&   --    &  -- &  4.16   & 1.10  &   1.37   &   \citet{Mar11}\\
NGC6715 &  $-$1.73 & 54   & $-$1.39 & 46  &   --    &  -- &  14.1   & 4.93  &  18.78   &  \citet{Mar19} \\
NGC6934	&  $-$1.55 & 93.3 & $-$1.34 & 6.7 &   --    &  -- &  1.17   & 0.33  &   0.1798 &  This work \\
NGC7089 &  $-$1.68 & 96   & $-$1.51 & 3.0 & $-$1.03 &   1 &  5.82   & 0.28  &   0.7805 &   \citet{Yong14}\\
NGC5139 &  $-$1.85 & 36   & $-$1.50 &  50 & $-$1.00 &   14&  35.5   & 15.11 &   96.52  & This work \\\hline
\end{tabular}
\end{table*}

\clearpage

\startlongtable
\begin{deluxetable}{lccccccc}
\tablewidth{10pt}
\tablecaption{Sensitivity of the derived abundances from UVES spectra to the
  uncertainties in atmospheric parameters
  (\teff/\logg/\vmicro/[A/H]=$\pm$50~K/$\pm$0.15~dex/$\pm$0.20~\kmsec/$\pm$0.10~dex), 
  and uncertainties due to the errors in the EWs measurements or in
  the $\chi$-square fitting procedure. For reference, we also list the
  variations due to a change in \teff\ by $\pm$100~K. We reported the total internal
  uncertainty ($\sigma_{\rm total}$) obtained by the quadratic sum of
  all the contributers to the error.}\label{tab:err} 
\tablehead{
      &\colhead{$\Delta$\teff} & \colhead{$\Delta$\teff}&\colhead{$\Delta$\logg}&\colhead{$\Delta$\vmicro} &\colhead{$\Delta$[A/H]} &\colhead{$\sigma_{\rm EWs/fit}$}&\colhead{$\sigma_{\rm total}$}\\  
      &\colhead{$\pm$100~K}    & \colhead{$\pm$50~K}    & \colhead{$\pm$0.15}   &\colhead{$\pm$0.20~\kmsec}& \colhead{$\pm$0.10~dex}&                             &                   }
\startdata
$\rm {[O/Fe]}$           & $\pm$0.02   &$\pm$0.01    & $\pm$0.07  & $\pm$0.00  & $\pm$0.03 & $\pm$0.05       & $\pm$0.09  \\
$\rm {[Na/Fe]}$          & $\mp$0.02   &$\mp$0.01    & $\mp$0.02  & $\pm$0.04  & $\pm$0.03 & $\pm$0.04       & $\pm$0.07 \\
$\rm {[Mg/Fe]}$          & $\mp$0.02   &$\mp$0.01    & $\mp$0.03  & $\mp$0.01  & $\mp$0.00 & $\pm$0.08       & $\pm$0.09   \\
$\rm {[Al/Fe]}$          & $\pm$0.07   &$\pm$0.05    & $\mp$0.01  & $\pm$0.00  & $\pm$0.01 & $\pm$0.08       & $\pm$0.12 \\
$\rm {[Si/Fe]}$          & $\mp$0.09   &$\mp$0.06    & $\pm$0.01  & $\pm$0.06  & $\pm$0.01 & $\pm$0.04       & $\pm$0.09    \\
$\rm {[Ca/Fe]}$          & $\pm$0.02   &$\pm$0.01    & $\mp$0.02  & $\mp$0.02  & $\mp$0.01 & $\mp$0.02       & $\pm$0.04 \\
$\rm {[Sc/Fe]}$\,{\sc ii}& $\pm$0.07   &$\pm$0.04    & $\mp$0.01  & $\pm$0.01  & $\mp$0.01 & $\pm$0.07       & $\pm$0.08   \\
$\rm {[Ti/Fe]}$\,{\sc i} & $\pm$0.09   &$\pm$0.05    & $\mp$0.02  & $\pm$0.01  & $\mp$0.01 & $\pm$0.03       & $\pm$0.06    \\
$\rm {[Ti/Fe]}$\,{\sc ii}& $\pm$0.05   &$\pm$0.03    & $\mp$0.02  & $\mp$0.02  & $\mp$0.02 & $\pm$0.06       & $\pm$0.07    \\
$\rm {[V/Fe]}$           & $\pm$0.10   &$\pm$0.05    & $\mp$0.01  & $\pm$0.03  & $\pm$0.03 & $\pm$0.06       & $\pm$0.08    \\
$\rm {[Cr/Fe]}$\,{\sc i} & $\pm$0.07   &$\pm$0.03    & $\mp$0.02  & $\mp$0.01  & $\mp$0.02 & $\mp$0.06       & $\pm$0.07    \\
$\rm {[Cr/Fe]}$\,{\sc ii}& $\pm$0.01   &$\pm$0.01    & $\mp$0.01  & $\pm$0.02  & $\mp$0.02 & $\pm$0.10       & $\pm$0.10    \\
$\rm {[Mn/Fe]}$          & $\pm$0.15   &$\pm$0.07    & $\mp$0.00  & $\mp$0.01  & $\mp$0.02 & $\pm$0.05       & $\pm$0.09  \\
$\rm {[Fe/H]}$\,{\sc i}  & $\pm$0.10   &$\pm$0.05    & $\pm$0.01  & $\mp$0.07  & $\pm$0.00 & $\pm$0.01       & $\pm$0.09   \\
$\rm {[Fe/H]}$\,{\sc ii} & $\mp$0.09   &$\mp$0.05    & $\pm$0.07  & $\mp$0.05  & $\pm$0.04 & $\pm$0.03       & $\pm$0.11   \\
$\rm {[Co/Fe]}$          & $\pm$0.00   &$\mp$0.00    & $\mp$0.00  & $\pm$0.05  & $\pm$0.01 & $\pm$0.10       & $\pm$0.11    \\
$\rm {[Ni/Fe]}$          & $\mp$0.03   &$\mp$0.02    & $\pm$0.01  & $\pm$0.04  & $\pm$0.01 & $\pm$0.03       & $\pm$0.05    \\
$\rm {[Cu/Fe]}$          & $\pm$0.11   &$\pm$0.05    & $\mp$0.02  & $\mp$0.04  & $\mp$0.03 & $\pm$0.11       & $\pm$0.13    \\
$\rm {[Zn/Fe]}$          & $\mp$0.12   &$\mp$0.06    & $\pm$0.00  & $\pm$0.05  & $\pm$0.01 & $\pm$0.15       & $\pm$0.17    \\
$\rm {[Y/Fe]}$\,{\sc ii} & $\pm$0.08   &$\pm$0.05    & $\mp$0.01  & $\mp$0.01  & $\mp$0.01 & $\pm$0.15       & $\pm$0.16    \\
$\rm {[Zr/Fe]}$\,{\sc ii}& $\mp$0.15   &$\mp$0.08    & $\pm$0.12  & $\mp$0.02  & $\pm$0.05 & $\pm$0.10       & $\pm$0.18    \\
$\rm {[Ba/Fe]}$\,{\sc ii}& $\pm$0.03   &$\pm$0.02    & $\pm$0.07  & $\mp$0.14  & $\pm$0.03 & $\pm$0.10       & $\pm$0.19    \\
$\rm {[La/Fe]}$\,{\sc ii}& $\pm$0.02   &$\pm$0.01    & $\pm$0.05  & $\mp$0.01  & $\pm$0.03 & $\pm$0.02       & $\pm$0.06    \\
$\rm {[Ce/Fe]}$\,{\sc ii}& $\pm$0.01   &$\pm$0.00    & $\pm$0.06  & $\mp$0.01  & $\pm$0.04 & $\pm$0.10       & $\pm$0.12    \\
$\rm {[Pr/Fe]}$\,{\sc ii}& $\pm$0.02   &$\pm$0.01    & $\pm$0.06  & $\pm$0.00  & $\pm$0.05 & $\pm$0.10       & $\pm$0.13    \\
$\rm {[Nd/Fe]}$\,{\sc ii}& $\pm$0.12   &$\pm$0.06    & $\mp$0.01  & $\mp$0.02  & $\mp$0.01 & $\pm$0.06       & $\pm$0.09    \\
$\rm {[Eu/Fe]}$\,{\sc ii}& $\mp$0.01   &$\pm$0.00    & $\pm$0.06  & $\pm$0.00  & $\pm$0.04 & $\pm$0.09       & $\pm$0.12   \\
\hline
\enddata
\end{deluxetable}

\end{document}